\documentclass{aa}

\def\teff{{\rm\,T_{eff}}}
\def\eg{{e.g.,{} }}
\def\CaHK{$\mathrm{Ca}\,\mathrm{H\&K}$\xspace}
\newcommand{\logg}{\ensuremath{\log g}\xspace}

\usepackage{graphicx,txfonts}
\usepackage{hyperref}
\hypersetup{
  colorlinks   = true,    % colors links instead of ugly boxes
  urlcolor     = blue,    % color for external hyperlinks
  linkcolor    = blue,    % color of internal links
  citecolor    = blue      % color of citations
}

\begin{document}

\title{The Pristine Inner Galaxy Survey (PIGS)}
\subtitle{XI: Revealing the chemical evolution of the interacting Sagittarius dwarf galaxy}

\author{
Sara Vitali\inst{\ref{UDP}, \ref{ERIS}}\thanks{analysis of data taken as part of ESO program ID 109.234B.001} \and Alvaro Rojas-Arriagada\inst{\ref{USACH}, \ref{ERIS},\ref{MIA}, \ref{CIR}} \and Paula Jofré \inst{\ref{UDP}, \ref{ERIS}} \and Federico Sestito\inst{\ref{CAR}} \and Joshua Povick\inst{\ref{UDP}, \ref{ERIS}}\and Vanessa Hill \inst{\ref{OBS_n}} \and Emma Fernández-Alvar \inst{\ref{IAC_1}, \ref{IAC_2}} \and Anke Ardern-Arentsen \inst{\ref{Cambr}}\and Pascale Jablonka \inst{\ref{EPFL}} \and Nicolas F. Martin \inst{\ref{CNRS},\ref{MAX}}\and Else Starkenburg  \inst{\ref{Kapteyn}} \and David Aguado\inst{\ref{IAC_1},\ref{IAC_2}}
}

\institute{
Instituto de Estudios Astrof\'isicos, Facultad de Ingenier\'ia y Ciencias, Universidad Diego Portales, Av. Ej\'ercito Libertador 441, Santiago, Chile \label{UDP}, \email{sara.vitali@mail.udp.cl}
 \and
 Millenium Nucleus ERIS \label{ERIS}
 \and
 Millenium Institute of Astrophysics, Av. Vicuña Mackenna 4860, 82-0436 Macul, Santiago, Chile \label{MIA}
 \and
 Departamento de F\'isica, Universidad de Santiago de Chile, Av. Victor Jara 3659, Santiago, Chile \label{USACH}
 \and
 Center for Interdisciplinary Research in Astrophysics and Space Exploration (CIRAS), Universidad de Santiago de Chile, Santiago,
 Chile \label{CIR}
 \and
 Centre for Astrophysics Research, Department of Physics, Astronomy and Mathematics, University of Hertfordshire, Hatfield, AL10 9AB, UK \label{CAR}
 \and
 Université Côte d’Azur, Observatoire de la Côte d’Azur, CNRS, Laboratoire Lagrange, Nice, France \label{OBS_n}
 \and
 Instituto de Astrofísica de Canarias (IAC), Vía Láctea, 38205 La Laguna, Tenerife, Spain \label{IAC_1}
 \and 
 Universidad de La Laguna, Dept. Astrofísica, 38206 La Laguna, Tenerife, Spain \label{IAC_2}
 \and
 Institute of Astronomy, University of Cambridge, Madingley Road, Cambridge CB3 0HA, UK \label{Cambr}
 \and
 Laboratoire d’astrophysique, École Polytechnique Fédérale de Lausanne (EPFL), CH-1290 Versoix, Switzerland \label{EPFL}
 \and
 Universit\'e de Strasbourg, CNRS, Observatoire astronomique de Strasbourg, UMR 7550, F-67000 Strasbourg, France \label{CNRS}
 \and
 Max-Planck-Institut f\"{u}r Astronomie, K\"{o}nigstuhl 17, D-69117 Heidelberg, Germany \label{MAX}
 \and
 Kapteyn Astronomical Institute, University of Groningen, Landleven 12, 9747 AD Groningen, The Netherlands \label{Kapteyn}
 }
\date{Accepted May~14,~2025; received December~9,~2024}

\abstract
 {The Sagittarius dwarf spheroidal galaxy (Sgr dSph) is a satellite orbiting the Milky Way that has experienced multiple stripping events due to tidal interactions with our Galaxy. Its accretion history has led to a distinct stellar overdensity, which is the remnant of the core of the progenitor.}
  % aims heading (mandatory)
 {We present a complete chemical analysis of 111 giant stars in the core of Sgr dSph to investigate the chemical evolution and enrichment history of this satellite. }
  % methods heading (mandatory)
   {Employing the metallicity-sensitive Ca H\&K photometry from the Pristine Inner Galaxy Survey, we selected stars that span a wide metallicity range and obtained high-resolution spectra with the ESO FLAMES/GIRAFFE multiobject  spectrograph. For the stellar sample covering $-2.13 < \rm{[Fe/H] < -0.35}$, we derived abundances for up to\,14 chemical elements with average uncertainties of $\sim 0.09$ dex and a set of stellar ages that allowed us to build an age-metallicity relation (AMR) for the entire sample.}
  % results heading (mandatory)
   {With the most comprehensive set of chemical species measured for the core of Sgr (Na, Mg, Al, Si, Ca, Sc, Ti, V, Cr, Co, Ba, La, and Eu), we studied several [X/Fe] ratios. Most trends align closely with Galactic chemical trends, but notable differences emerge in the heavy $n$-capture elements, which offer independent insights into the star formation history of a stellar population.} 
   {The deficiency in $\alpha$ elements  with respect to the Milky Way suggests a slower, less efficient early star formation history, similar to other massive satellites.
   $S$-process element patterns indicate significant enrichment from Asymptotic giant branch stars over time. The AMR and chemical ratios point to an extended star formation history, with a rapid early phase in the first  gigayears, followed by declining activity and later star-forming episodes. These findings are consistent with Sgr hosting multiple stellar populations, from young ($\sim 4$ Gyr) to old, metal-poor stars ($\sim 10$ Gyr).}


\maketitle

\section{Introduction}

The Lambda cold dark matter ($\Lambda$CDM) model provides the most up-to-date understanding of galaxy formation. It suggests that galaxies form through the merging and subsequent accretion of smaller systems. This hierarchical model posits that galaxies enclose dust, gas, and stars, all embedded within massive dark matter halos \citep{1978Searle,1978White,1991White}. In this context, the Milky Way (MW) serves as an ideal laboratory, as its spatial resolution allows for the study of individual stars and the acquisition of detailed information to explore its star formation history. At smaller scales, the accretion events that built up our Galaxy can be better understood by studying the surrounding dwarf galaxies that comprise the majority of the MW's satellite populations \citep[e.g.,][] {2004Venn,2009Tolstoy,2016BlandHawthorn}.

Of such galaxies, one of the biggest and most luminous known around the MW is the Sagittarius (Sgr) dwarf spheroidal (dSph) galaxy, discovered by \cite{1994Ibata}. With an estimated total mass of $\sim 4\times 10^{8} \rm{M_{\odot}}$ \citep{2012McConnachie,Vasiliev20} and a luminosity of $M \sim -13.5$ \citep{2012McConnachie}, Sgr is an ideal "test case" for investigating chemical evolution theories within the framework of hierarchical accretion and learn about the early evolution of the MW. In addition, Sgr is currently undergoing a merger with the MW, having experienced its first infall approximately 5 Gyr ago \citep{2020Ruiz}. The long tidal interaction with our Galaxy has profoundly shaped Sgr, leaving behind a remnant core located on the opposite side of the Galactic center at a  heliocentric distance $d \sim 26.5$ kpc \citep{1994Ibata, 2003Majewski,2010Law,Vasiliev20}, along with two wide stellar streams.

Numerous studies, both photometric \citep[e.g.,][]{1999BellazziniI, 2000Layden, Siegel07, Vitali22} and spectroscopic \citep[e.g.,][]{2000Bonifacio, 2005Monaco, 2013McWilliam, 2017Hasselquist,2018Hansen,2021Hasselquist,2024Sestito, 2024Sestitocar} that have targeted the Sgr core in recent years, have revealed an extended star formation history (SFH) heavily influenced by gravitational interactions with the MW. During Sgr's orbital passages around the MW, multiple star formation episodes occurred, leading to the formation of distinct stellar populations: a predominant intermediate-age population (4-8 Gyr, -0.6 $\lesssim \rm{[Fe/H]} \lesssim-0.4$), a young population ($\sim 2.5$ Gyr and [Fe/H]$\sim 0$), and a very young population of a few megayears with supersolar metallicities (age $<2$ Gyr and [Fe/H]$\sim +0.5$) \citep{layden1997sgr,Siegel07,1999Bellazzini}. These populations are hosted in the core of Sgr alongside the older, more metal-poor population (age $>10$ Gyr, [Fe/H]$\lesssim-1.0$). Part of the most metal-poor population has been gradually stripped away from the core of Sgr during its history of interaction with the MW, and deposited in the two long stellar streams that wrap around it \citep{2001Ibata, 2003Majewski}. These streams are known to be $\sim 1$ dex more metal poor than the core \citep{2015deboer}.

Despite a fair understanding of the evolution of this dissolving system, its SFH and how its evolution has been affected by interactions with the MW remain open questions. To address this topic, the exquisite data releases of \textit{Gaia} \citep{2016Gaia,2021Gaia,2023Gaia}, which have collected astrometric and photometric data for 1.8 billion sources in the MW and its satellites, can be efficiently combined with photometric surveys such as \textit{Pristine} \citep{2017Starkenburg, 2023Martin}, which is specifically dedicated to tracing the history of the MW and its satellites through the study of metal-poor stars. This latter survey, primarily focused on the MW halo, has developed a particular focus on the inner Galactic region thanks to the \textit{Pristine} Inner Galaxy Survey \citep[PIGS,][]{2020Arentsena, 2024Arentsen}. Both surveys rely on a narrow-band filter centered around the CaH\&K region to identify metal-poor candidates. Their efficiency has led to numerous studies focused on both our Galaxy and the Sgr region \citep{Arentsen20b,Vitali22, 2024Sestito, 2024Sestitocar}, which is included in the coverage of PIGS. Using this dataset, \cite{Vitali22} performed a photometric analysis and characterized the metallicity distribution of approximately 50,000 Sgr members, finding a negative photometric metallicity gradient extending up to 5.5 kpc along the main body of Sgr, indicative of an outside-in formation process. \cite{2024Sestito} chemically characterized 12 very metal-poor (VMP) stars spectroscopically, and provide new insights into the early evolution of Sgr by assembling the largest high-resolution study in the VMP regime and significantly expanding the existing sample of VMPs \citep{2018Hansen, 2019Chiti, 2020Chiti}. In a recent study, \cite{2024Sestitocar}extend the investigation of the ancient chemical evolution of Sgr by analyzing carbon abundances using low-/medium-resolution spectroscopic follow-up data from PIGS \citep{Arentsen20b, 2024Arentsen}. 

Despite these detailed analyses, a statistically robust characterization of Sgr's evolution needs to be extended to a broader metallicity range to extend beyond the early stages of this satellite. To achieve this, it is crucial to continue gathering precise and accurate measurements of various chemical species across a wider range of metallicities and for a larger sample of stars. For instance, constraining the trend of $\alpha$ elements with metallicity allows for a deeper understanding of the initial mass function and the SFH of a system \citep[e.g.,][]{tinsley1968,mcwilliam1997}. Simultaneously, obtaining additional measurements of heavier elements (atomic number $Z > 30$), which are produced in both the slow ($s$) and rapid ($r$) neutron-capture processes \citep[e.g.,][]{2004Travaglio,karakas2014dawes, 2014Bisterzo, limongi2018}, is crucial for probing the astrophysical sites of these processes, unraveling their distinct chemical signatures, and exploring the timescales associated with the SFH of the galaxy. By adding high-resolution chemical abundances to the already existing dataset on Sgr \citep[e.g.,][]{2013McWilliam,2005Monaco,2007Sbordone,2010Carretta,2018Hansen}, this analysis enables comparative studies of chemical evolution between Sgr and other MW satellites, such as the Sculptor and Fornax dSphs, the Large Magellanic Cloud, and the accreted Gaia-Enceladus/Sausage system.

In this work, we present a sample of 111 Sgr giant stars for which we obtained high-resolution spectra thanks to observations performed with the ESO FLAMES/GIRAFFE multiobject spectrograph \citep{2002Pasquini}. Through a comprehensive spectroscopic analysis, we intend to study the evolution of the Sgr system and compare it with other massive MW satellites that have different masses and SFHs. Additionally, to the best of our knowledge, this spectroscopic sample represents the largest optical high-resolution dataset ($R>19000$) with 13 elements derived in a homogeneous way. The target selection and data reduction are detailed in Section \ref{sect:data}. Section \ref{sect:spec} describes the spectroscopic analysis, while Section \ref{sect:discussion} discusses the main conclusions that can be drawn from the chemical abundance analysis regarding the chemical evolution of Sgr, comparing it with other known MW systems. The conclusions are summarized in Section \ref{sect:conclusion}. 

\begin{figure}[t]
\centering
 \includegraphics[width=\columnwidth]{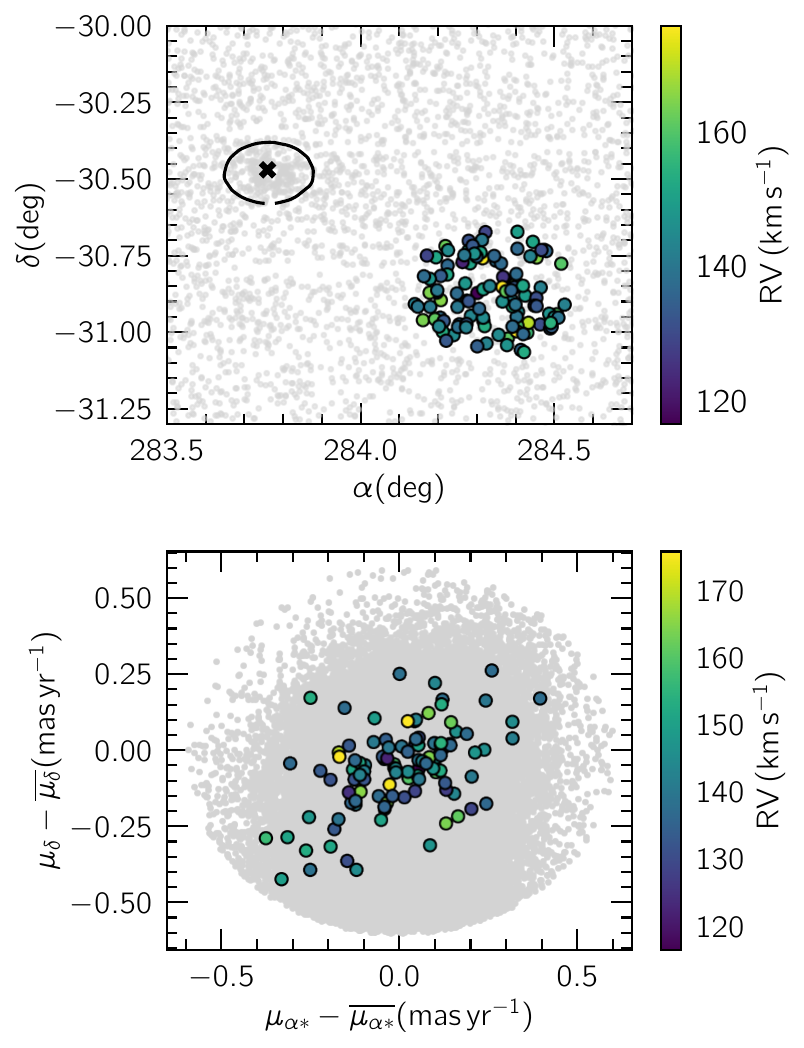} 
\caption{\small{Target selection. Top: Zoom in equatorial coordinates on the PIGS-Sgr photometric footprint after all the cuts (gray points are the Sgr candidate members). The black cross identifies the nuclear globular cluster M54, and the contour ellipse marks its spatial extent. The FLAMES targets are colored according to their radial velocities. Bottom: Proper motion space (with mean values subtracted) for the Sgr-PIGS selection and the FLAMES targets.}}
 \label{fig:tar}
\end{figure}
\section{Data} \label{sect:data}

\subsection{Target selection}\label{sect:targe}
The targets are located in the fields observed by PIGS \citep{2020Arentsena}, an extension of the \textit{Pristine} survey. A specific region of PIGS is focused on the Sagittarius system, resulting in the collection of metallicity-sensitive \CaHK photometry for the remaining core of the galaxy up to the onset of the galaxy's tidal stellar streams \citep{2020Arentsena,Vitali22}. 

From these photometric measurements, the very metal-poor (VMP) candidates ([Fe/H] < $-$2.0 dex) identified by PIGS were followed up using the AAOmega+2dF facility mounted on the Anglo-Australian Telescope \citep[AAT,][]{Saunders2004,2002Lewis,Sharp2006}. From the collected low-/medium-resolution spectra ($R \sim 1300 - 11\,000$), effective temperatures, surface gravities, metallicities, and carbon abundances were derived for $\approx$ 13000 VMP stars (see \citealt{2020Arentsena,2024Arentsen} for details). Four AAT pointings were dedicated to collecting spectra for about 400 members of the Sagittarius system (AAT-Sgr), which have already been used in various works (see for instance \citealt{Vitali22, 2024Sestito, 2024Sestitocar}). Overall, these spectroscopic follow-up efforts have demonstrated the power of PIGS photometry as a metallicity predictor.

For this work the Sgr targets were selected from the PIGS photometric catalog, as there is not corresponding AAT spectroscopy. Their properties are displayed in Fig.~\ref{fig:tar}. This central field (colored points) lies in the core of the galaxy but is far enough ($\sim 1^{\circ}$ in both RA and DEC) from the nuclear cluster M54/NGC 6715, marked by a black cross and an ellipse indicating its spatial extent, to exclude stars from the cluster in our selection. To include only Sgr members and exclude MW foreground stars (where the Sgr stars are shown as gray points in Fig.~\ref{fig:tar}) from the PIGS photometric catalog (hereafter the PIGS-Sgr), we applied the membership criterion based on \textit{Gaia} photometry and astrometry presented in \cite{Vitali22} and \cite{2024Sestitocar}. In addition to the quality cuts on \textit{Gaia} photometry and astrometry adopted in these previous studies, we applied a cut on the parallax and parallax error ($|\overline{\omega}|\leq 2\epsilon_{\overline{\omega}}$) to remove nearby MW stars. We limited stars taking into account the change in average proper motions along the body of Sg within a radius of 0.6 mas $\mathrm{yr^{-1}}$ \citep{Vasiliev20} as depicted on the bottom panel of Fig.~\ref{fig:tar}. Finally, to ensure a high signal-to-noise ratio (S/N) for the spectra in the observations, the limiting magnitude was set to $G = 17$. As shown in section \ref{sect:RV} the radial velocities of our targets fall within the typical range reported in the literature for Sgr, reinforcing the membership of these stars to the dwarf system. 

\begin{figure}[t]
\centering
 \includegraphics[width=\columnwidth]{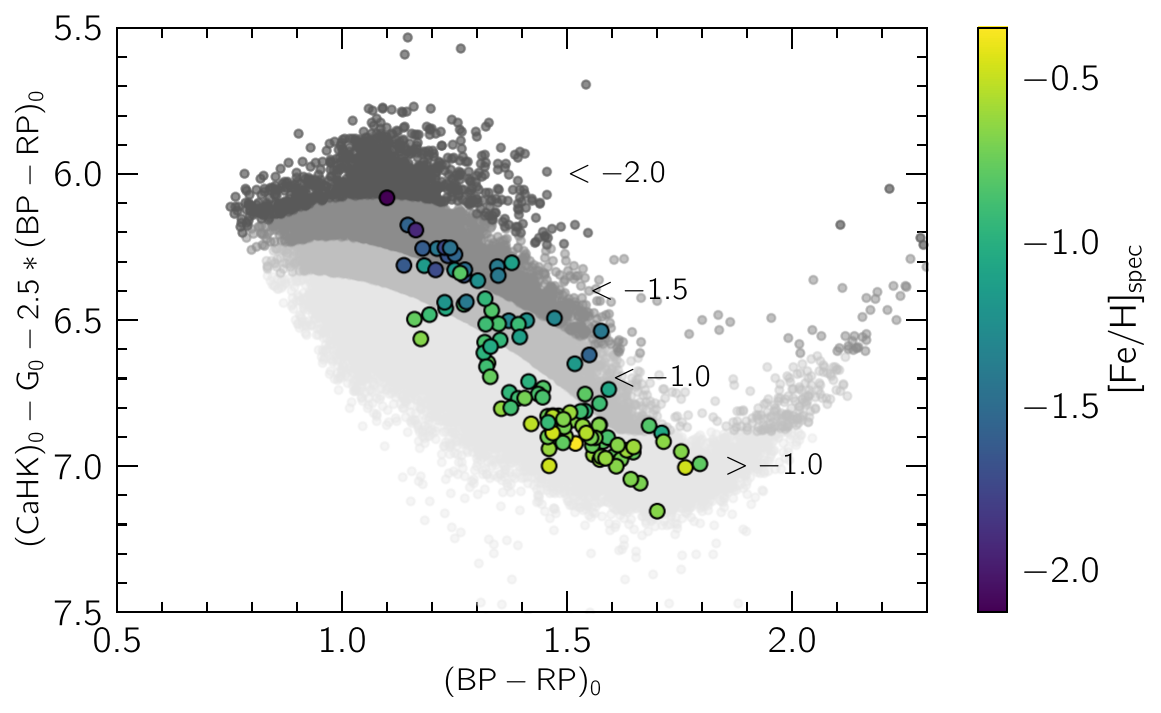} 
\caption{\small{Color-color diagram for the PIGS-Sgr sample. The four metallicity bins are identified with four different shades of gray and labeled according to their ranges. The FLAMES targets are overplotted as circles with black edges, colored by their spectroscopic metallicities.}}
 \label{fig:CCD}
\end{figure}

Among the selected PIGS-Sgr members, we used the PIGS photometry to build a stellar sample with an extended metallicity range. This approach allowed us to include representatives of the various stellar populations within the system for a comprehensive chemical study. To do so, we relied on the same method featured in \cite{Vitali22}. This strategy combined the spectroscopic training sample from the main \textit{Pristine} halo survey \citep{2017Starkenburg} with the \textit{Pristine} Ca H\&K catalog \citep{2023Martin} to derive iso-metallicity lines in the Ca H\&K-\textit{Gaia} color-color space. Consistent with the approach of \cite{Vitali22}, we also applied an extinction correction to the PIGS photometry. 

Limiting the selection to giant stars ($\logg < 3.98$, $\teff < 5700$ K and $-4.0 < \rm{[Fe/H]} < -0.5$) we obtained the diagram shown in Fig.~\ref{fig:CCD}. The colored areas are delimited by the iso-metallicity lines, which divide the PIGS-Sgr color-color space into four different metallicity bins (different gray points): [Fe/H] $< -2.0$; $-2.0 <$ [Fe/H] $< -1.5$; $-1.5 <$ [Fe/H]  $< -1.0$; and $-1.0 < $ [Fe/H] $< -0.5$. As described in \cite{Vitali22} an offset ($\sim 0.52$ mag) was applied to the PIGS-Sgr photometry to align it with the training sample, which was calibrated against the main survey where the Ca H\&K photometry has a constant offset. This offset arises because, at the time, the "main" \textit{Pristine} fields were calibrated using a part of the Ca H\&K-SDSS color-color space populated mainly by nearby dwarfs \citep{2017Starkenburg}, for which the adopted absolute scale was different from that acting on the regions of the PIGS-Sgr photometry. The final Sgr candidates were selected from this color diagram after applying the offset, ensuring the inclusion of stars across various metallicity bins, while also considering the configuration and positional constraints of the FLAMES/GIRAFFE 130 instrument fibers, which collectively cover a field of view 25 arcmin in diameter (for more details see Sect.~\ref{sec:obs}). In fact, limitations such as avoiding placing fibers on too bright stars, selecting a sufficiently crowded field, and respecting the magnitude limit to ensure reasonable exposure times had to be balanced with the need to cover a reasonably extended metallicity range for our targets.

The final targets, hereafter PIGS-HR-Sgr, are displayed in the same color-color diagram (Fig.~\ref{fig:CCD}), color coded according to their spectroscopic metallicities (derived as explained in Section \ref{sect:metallicities}). Their positions on the diagram confirm the sensitivity of the PIGS photometry to metallicity and the accuracy of our measurements, as the majority of stars fall into the correct metallicity bins based on their spectroscopic [Fe/H] values. However, there is an exception for a group of stars with $\mathrm{[Fe/H]_{spec}} \sim -1.3/ -1.4$, that the Ca H\&K photometry classifies as more MP. For these stars, it is possible that the Ca H\&K uncertainties (0.2-0.8 mag) played a role in their selection. On the other hand, considering the spectroscopic $\sigma_{\mathrm{[Fe/H]}}$, which are around $\sim 0.1$ dex, these stars might be placed in the adjacent metallicity bin. Finally, another factor to consider is that Sgr has lower $\rm{[\alpha/Fe]}$ than the training sample from \textit{Pristine}, which is used to define the iso-metallicity lines. Indeed, the training sample is representative of the MW, with higher $\rm{[\alpha/Fe]}$ than is typical for Sgr \citep{2017Hasselquist}, as already discussed in \cite{Vitali22}.
\begin{table}[!t]
\caption{Observing setup used during the observation.}
    \centering
    \begin{tabular}{c c c c c c} 
    \hline \hline
    \small
Filter  & $\mathrm{\lambda_{in}}$ - $\mathrm{\lambda_{fin}}$ (nm) & $R$ & $<\rm{S/N}> $&$\mathrm{Exp_{time}}$ (h) \\
        \hline \hline
            \small
HR8 & 491.7 - 516.3 & 23500 & 40 &  6.67 \\
HR11 & 559.7 - 584 & 29500 & 33& 3.42 \\
HR15N & 647 - 679 & 19200 & 53 & 1.65  \\
         \hline
    \end{tabular}
\tablefoot{\small{Wavelength coverage, resolution, average signal-to-noise ratio, and exposure time of the GIRAFFE setups used for this work. The signal-to-noise ratio is given per pixel (1 pixel $= 0.005$ nm).}}
    \label{tab:obser}
\end{table}

\subsection{Observations} \label{sec:obs}

The observations were carried out between April-May 2022 using the ESO FLAMES/GIRAFFE multiobject spectrograph \citep{2002Pasquini} mounted on the Very Large Telescope (VLT). For the selected field, the combination of HR8, HR11 and HR15N setups (see Table \ref{tab:obser}) were employed to observe 111 giant stars with  a total wavelength coverage of 80.9 nm. Their spectra were taken in MEDUSA mode, and their S/N per pixel (1 pixel $= 0.005$ nm) span from 20-80 for HR8, 20-66 for HR11 and 29-99 for HR15. Table \ref{tab:obser} reports the mean S/N for each setup, while the S/N values for the individual stars are provided in a complete table available at the CDS, along with comprehensive target information. 

\subsection{Spectra preparation}

We retrieved the reduced data from the GIRAFFE ESO Phase 3 Data Stream reduction\footnote{\url{http://www.eso.org/sci/observing/phase3/data_streams.html}}.The data were reduced, extracted, cleaned for cosmics and wavelength calibrated thanks to the GIRAFFE Base Line Data Reduction Software (BLDRS)\footnote{\url{http://girbldrs.sourceforge.net}}. 
\begin{figure}[t]
\centering
 \includegraphics[width=\columnwidth]{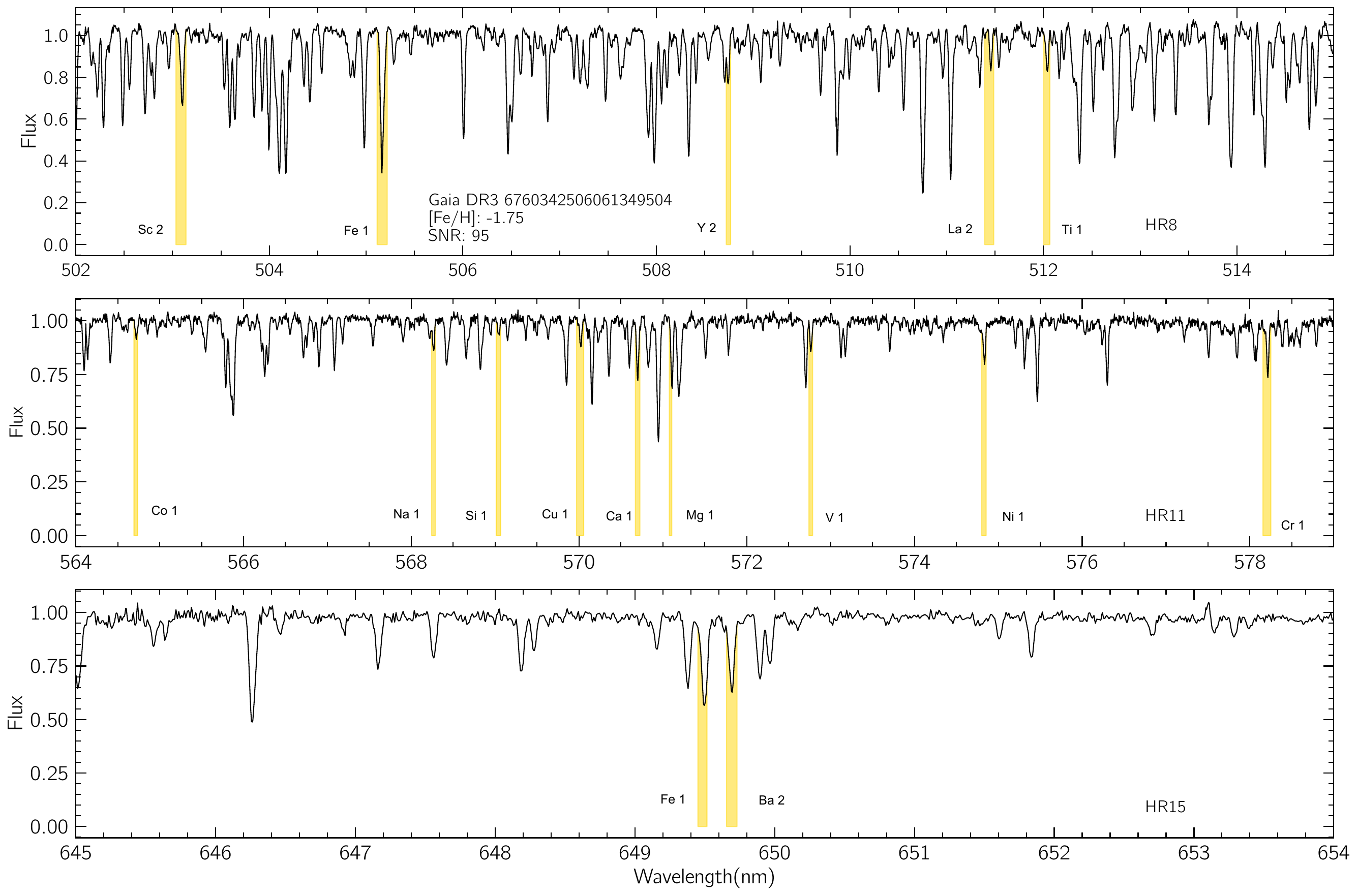} 
\caption{\small{Examples of spectral lines from a FLAMES spectrum in three different setups. Top: The HR8 setup includes Sc II (503.1 nm), Fe I (505. nm), Y II (508.7 nm), and Ti I lines (512.2 nm), as well as the La II line at 511.4 nm, all marked with colored, shaded regions. Central: For the HR11 filter, we display the following lines: Co I (564.7 mn), Na I (568.2 nm), Si I (569.04 nm), Cu I (570.02 nm), Ca I (570.1 nm), one line of the Mg I \textit{b} Triplet region (571.1 nm), V I (570.1 nm), Ni I (574.8 nm), and Cr I (578.1 nm). Bottom: For the HR15 setup, the colored areas highlight the Ba line (649.6 nm) plus an adjacent Fe line (649.6 nm).}}
 \label{fig:spectra}
\end{figure}

We built an automatic multi-step pipeline that processes all of the individual exposures of the spectra to produce a final set of co-added spectra for each star. First, instrument signatures and sky emission lines are detected and removed, and the affected pixels are flagged, as their S/N is lower and some degree of residuals may persist. Next, it performs continuum normalization by iteratively (10 iterations) fitting a cubic spline while rejecting pixels using the standard deviation of the residuals as a metric. Finally, the multiple exposures of a given star are resampled to a common wavelength and co-added using mean. In the combination, pixels flagged in previous steps are ignored. Examples of reduced and normalized spectra are reported in Fig.~\ref{fig:spectra}. The panels display one target observed with three different grating setups. These examples have been reduced, normalized, and velocity-corrected. The corresponding S/N and [Fe/H] values are indicated in the top figure, along with examples of absorption lines marked in yellow.

\section{Spectroscopic analysis} \label{sect:spec}

\subsection{Radial velocities}\label{sect:RV}

The radial velocities (RVs) were computed using a custom-created grid of synthetic models as templates over a range of atmospheric parameters and metallicity representative of the observed sample. For each spectrum, a first cross-correlation is performed with a randomly selected template to determine a preliminary RV value. The corrected observed spectrum is then compared to the whole grid, and the template with the lower chi-square is selected to infer the final velocity value. When calculating the cross-correlation function (CCF), the algorithm refines the maximum value by fitting a Gaussian curve. It is worth mentioning that, when computing the individual abundances, a small correction in RV was allowed to the spectrum in the rest frame to improve the fit to the specific lines used to measure the single abundances. Moreover, the RVs were computed separately for each setup, with a mean dispersion of 0.6 $\mathrm{km,s^{-1}}$, which falls within the average RV uncertainties of the order of $\sim$ 0.9-1 $\mathrm{km,s^{-1}}$ (see Appendix~\ref{sec:append_rv}), slightly higher than those of the Sgr candidate sample selected from APOGEE DR17 \citep{APOGEEDR17}, which are on the order of $\sim 0.6\,\mathrm{km\,s^{-1}}$.
\begin{figure}[t]
\centering
 \includegraphics[width=0.8\columnwidth]{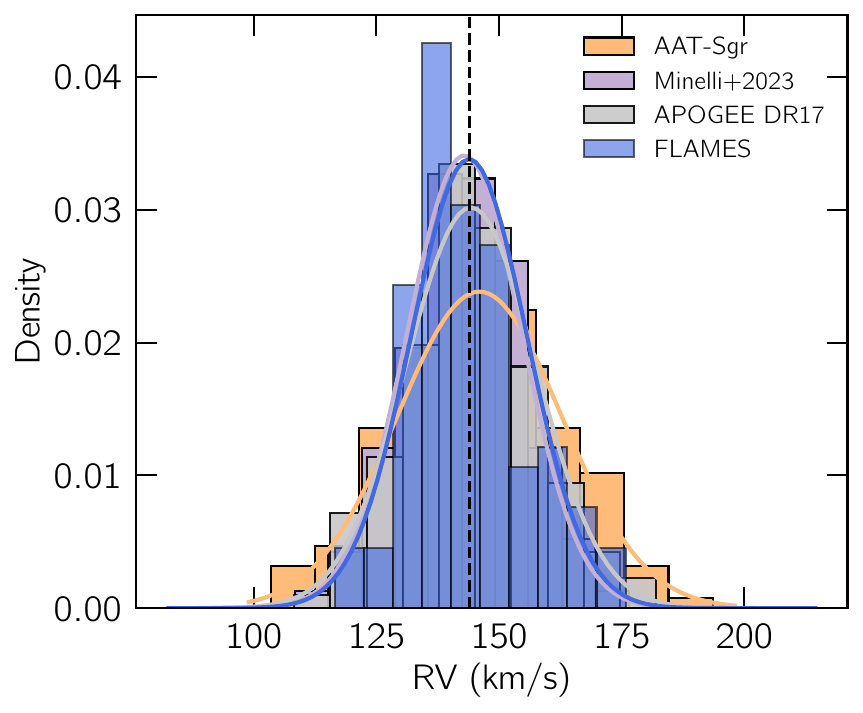} 
\caption{\small{Normalized distribution of RVs for Sgr samples. The target stars presented in this work appear in blue. The Sgr selections from \cite{2024Sestitocar}, \cite{2023Minelli} and \cite{APOGEEDR17} are colored according to the legend. The continuous lines represent KDE curves, colored to match the corresponding Gaussian distributions. The average RV values occur at $143.9\,\mathrm{km\,s^{-1}} $ marked with the vertical dashed line.}}
 \label{fig:RVhisto}
\end{figure}
\begin{figure*}[t]
\centering
 \includegraphics[width=\textwidth]{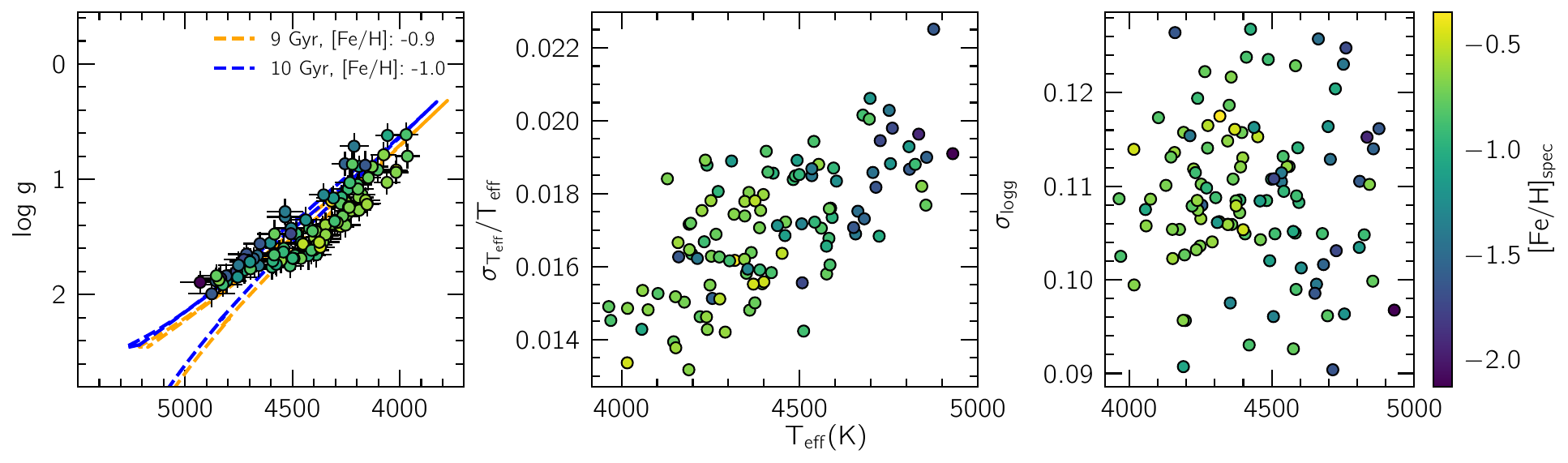} 
\caption{\small{Kiel diagram and parameter uncertainties of our stellar sample. Left: Kiel diagram with the parameters derived through photometry. All points are color coded by their spectroscopic metallicities. Two examples of BaSTI isochrones are shown and zoomed in around the evolutionary stage of our giant stars. Central and right: The $\teff$ relative uncertainties and $\logg$ absolute uncertainties vs. effective temperatures for the entire sample following the same color coding as the Kiel diagram.}}
 \label{fig:AP}
\end{figure*}

The final distribution of the Sgr RVs is illustrated in Fig.~\ref{fig:RVhisto}, which shows the RV values averaged over the three setups. Our FLAMES targets (in blue) are compared with the AAT-Sgr sample (in orange), the FLAMES high-resolution sample (light purple) from \cite{2023Minelli}, a selection of Sgr members from APOGEE DR17 (in gray), following the same selection (in coordinates: $b < -10^{\circ}$, proper motion space and radial velocity) criteria adopted in this work. These latter fall within the Sgr velocity distribution found in other studies, specifically 100-200 $\mathrm{km\,s^{-1}}$. The FLAMES distribution skews toward lower RVs, possibly due to the observations being concentrated on a specific, localized field. Adopting the same method described in \cite{2024Sestitocar}, we inferred the systemic RV and RV dispersion using a Monte Carlo Markov Chain (MCMC) approach. A step function was employed as the prior probability distribution, considering both the system's velocity dispersion and the uncertainties associated with individual RV measurements. By applying the same limits for the RV and its dispersion (i.e., $90 \leq \rm{RV} \leq 230 \mathrm{km,s^{-1}}$ and  $\rm{\sigma_{RV}} \leq 40 \mathrm{km,s^{-1}}$), we obtained a systemic $\langle \rm{RV} \rangle$ of $\sim 143.9\pm1.1 \,\mathrm{km\,s^{-1}} $ and a velocity dispersion $\sigma_{\rm{RV}}$ of $\sim 11.93\pm 0.8\,\mathrm{km\,s^{-1}} $, finding consistency with the values in the literature (\eg \citealt{1994Ibata,2021Delpino,2023Minelli, 2024Sestitocar, 2024An}). Specifically, examining the systematic RV values and corresponding dispersions from \cite{2024Sestitocar}, who divided their Sgr samples based on metallicity and on elliptical distance from the Sgr's center, our values are more closely aligned with those from their MR population ([Fe/H] $>-0.6$), in particular with both their total MR sample and the one at outer distances, defined as projected elliptical
distances $\geq 0.25 \rm{r_{h}}$. Given that the average elliptical distance for our FLAMES field is $\approx 0.34 \rm{r_{h}}$, computed adopting the same values (coordinates of the system center, ellipticity, half-light radius, and position angle) as in \cite{Vitali22} and \cite{2024Sestitocar}, our findings remain consistent with the outer MR group identified by \cite{2024Sestitocar}. In the case of \cite{2023Minelli}, the similarity is found with their total population rather than the values for their MP and MR samples, which can be attributed to the different metallicity cuts employed in these works.

\subsection{Atmospheric parameters}\label{sect:AP}

We followed the methodology described in \cite{2023Sestito} who employed an iterative process between the color-temperature relation from \cite{Mucciarelli21} and the Stefan-Boltzmann equation to determine the atmospheric parameters. This strategy has proven effective in reliably deriving stellar parameters when the wavelength coverage is limited, provided the photometric quality is comparable to that of the spectra \citep{2020Karovicova, 2020Mucciarelli,2023Sestito}. Besides, this approach is less affected by non-Local Thermodynamic Equilibrium (NLTE) effects \citep{2013Frebel, 2020Ezzeddine}. 

The sets of effective temperatures and surface gravities were determined through an iterative process, consisting of the following steps: $\teff$ was derived using the color-temperature relation with the line-by-line spectroscopic metallicities used as input (described in Sect.~\ref{sect:metallicities}). The surface gravity was calculated with the Stefan-Boltzmann equation \citep[e.g.,][]{Venn17}. This latter step requires temperatures (inferred in the previous step), the heliocentric distances, \textit{Gaia} magnitudes and bolometric corrections \citep{2018Andrae}. The necessary extinction correction was done with the 2D dust map from \cite{Schlegel98} and updated by \cite{Schlafly11}. The {\it Gaia} magnitudes were de-reddened using the extinction coefficients $\rm A_V/E(B-V)= 3.1$ \citep{Schultz75} and $\rm A_G/A_V = 0.85926$, $\rm A_{BP} /A_V = 1.06794$, $\rm A_{RP} /A_V = 0.65199$ from \cite{Marigo08,Evans18}. Absolute magnitudes were derived assuming a fixed distance of $26.5$ kpc for the Sgr stars \citep{Vasiliev20}. The errors on the distance were assumed to be 5 kpc, which should reasonably incorporate the physical thickness of the Sgr as it is estimated to range from $0.7$ to $2.6$ kpc \citep{Vasiliev20}. The masses computing the surface gravities were estimated assuming a uniform distribution between $0.5$ and $0.8 \rm{M_{\odot}}$ as a prior. These mass limits should be reasonable, as the majority of the stars are expected to be low-mass with ages of $\gtrsim 4$ Gyr. To verify the robustness of this assumption, we repeated these steps and recomputed the atmospheric parameters using a modified mass range of $0.6$–$1 \,\rm{M_{\odot}}$. The resulting differences in the surface gravities remained within the error bars. 

For assessing the uncertainties on the measured stellar parameters we employed a Monte Carlo simulation repeated 1000 times on the input parameters. 
This step accounts for uncertainties in \textit{Gaia} magnitudes, bolometric corrections, the heliocentric distance of Sgr (26.5 kpc), reddening (computed using the aforementioned extinction coefficients), and spectroscopic input metallicities, which are perturbed assuming a Gaussian distribution. The input mass values, however, are assumed to follow a flat distribution. The final mean uncertainties on $\teff$ is $75$ K, while on $\logg$ is 0.11 dex. 

As mentioned above, the color-temperature relation requires metallicities as input. These initial estimates were initially derived using the public spectroscopic software \texttt{iSpec} \citep{2014Blanco, 2019Blanco}. At this stage, metallicities, microturbulence velocity ($\mathrm{v_{mic}}$) and broadening parameters (resolution $R$ and macroturbulence velocity $\mathrm{v_{mac}}$) were inferred with \texttt{iSpec} via spectral synthesis. We adopted the radiative transfer code Turbospectrum v19.1 \citep{1998A&A...330.1109A,2012ascl.soft05004P}, in local thermodynamic equilibrium, and the one-dimensional spherical MARCS model atmospheres \citep{2008A&A...486..951G}. Atomic data were taken from the line-list designed for the \textit{Gaia}-ESO Survey (GES, \citealt{2021Heiter}). The $\mathrm{v_{mac}}$ was estimated using an empirical relation from the \textit{Gaia}-ESO survey. Building on these results, we then computed the photometric $\teff$ and $\logg$. Our final parameters are thus those that combine both photometric effective temperatures, surface gravities, and spectroscopic $\mathrm{v_{mic}}$ and Fe values. 
\begin{table*}[] \caption{Target information.} \label{tab:AP}
    \begin{tabular}{l c c c c c c c c c}
    \hline \hline
        \multicolumn{10}{c}{\footnotesize Target information and spectroscopic parameters}\\   
    \hline \hline 
    \scriptsize  Gaia DR3 ID &\scriptsize        S/N&\scriptsize         RA($\deg$) & \scriptsize DEC($\deg$)&\scriptsize         $\rm{G_{mag}}$ &\scriptsize RV $(\mathrm{km s}^{-1})$ & \scriptsize $\teff~\mathrm{(K)}$& \scriptsize        $\log\,g~\mathrm{(dex)}$& \scriptsize    [Fe/H]& \scriptsize    $v_{\rm mic} \,(\mathrm{km s}^{-1})$ \\
    \tiny 6760361266479328000 & \scriptsize      43 & \scriptsize        284.431 & \scriptsize    -30.967 & \scriptsize   16.397 & \scriptsize    169.59 $\pm$ 0.49& \scriptsize   4481$\pm$ 82 & \scriptsize      1.47 $\pm$ 0.11  & \scriptsize      -0.92 $\pm$0.09 & \scriptsize   1.53 $\pm$ 0.07 \\
    \tiny 6760381500040022528 & \scriptsize      45 & \scriptsize        284.312 & \scriptsize -30.698 & \scriptsize     16.7227 & \scriptsize 140.58 $\pm$ 0.56& \scriptsize  4582 $\pm$ 76 & \scriptsize 1.67 $\pm$ 0.10  & \scriptsize      -0.90 $\pm$0.10 & \scriptsize    1.46 $\pm$ 0.07 \\
    \hline\hline
        %\multicolumn{4}{c}{\footnotesize Spectroscopic parameters}\\    \hline \hline  
    \end{tabular}
\tablefoot{Example of the target information table, including \textit{Gaia} identifiers, the average S/N from the three observed setups, coordinates, G magnitudes, and our derived stellar photospheric parameters and radial velocities for two sample target stars. The complete table is provided at the CDS.}    
\end{table*}

The Kiel diagram for the entire sample is shown in the left panel of  Fig.~\ref{fig:AP}, together with their relative uncertainties or absolute uncertainty in the case of $\log g$. The stars are all in the giant evolutionary stage and present a metallicity from $\approx -2.11$ to  $-0.49$ dex. To guide the eye, we compare our parameters with two BaSTI isochrones \citep{2018Hidalgo,2024Pietrinferni} selected with the closest values to the mean [Fe/H] value of $-0.96$ dex found with spectroscopy. The ages of 9 and 10 Gyr of the isochrones are compatible with intermediate-age and old MP populations, with metallicities of $\sim-1$, as found in the study by \cite{Siegel07}. Looking at the central and right panels of Fig.~\ref{fig:AP} it is noticeable that the relative uncertainties in temperature increase for hotter stars and for the more MP ones, reaching $\sim 2.2$\%. The precision in $\log g$ shows no direct correlation with $\teff$ as it is most influenced by reddening effects, which are homogeneous within the FLAMES field. These trends might impact the derivation of elemental abundances for the colder and hotter targets, as well as for the most MP ones. The complete set of atmospheric parameters is provided as supplementary material at the CDS, following the format of Table \ref{tab:AP}.

\subsection{Chemical abundances}\label{sect:chemistry}
\begin{table}[]
\caption{Line selection and final chemical abundances.}\label{tab:lines}
    \begin{tabular}{l c c c}
    \hline \hline
        \multicolumn{4}{c}{\footnotesize Line selection}\\ 
    \hline\hline
    \scriptsize Element&\scriptsize $\lambda \rm{(nm)}$& \scriptsize $\rm{E}_{low}$ (eV)&\scriptsize $\log\,gf$\\
    \scriptsize Fe I & \scriptsize 491.899 & \scriptsize 2.865&\scriptsize -0.342\\
    \scriptsize Mg I & \scriptsize 571.109 & \scriptsize 4.346&\scriptsize -1.724\\
    \hline\hline
        \multicolumn{4}{c}{\footnotesize Chemical abundances}\\ 
    \hline\hline
    \scriptsize star & \scriptsize Element& \scriptsize [X/Fe] & \scriptsize $\rm{N_{lines}}$\\
    \tiny 6760361266479328000 & \scriptsize Mg I & \scriptsize -0.82 $\pm$ 0.09 & \scriptsize 1 \\
    \tiny 6760381500040022528 & \scriptsize Ca I & \scriptsize -0.61 $\pm$ 0.15 & \scriptsize 7 \\
    \hline\hline
        \multicolumn{4}{c}{\footnotesize Solar abundances}\\ 
        \hline\hline
    \scriptsize star & \scriptsize Element& \scriptsize A(X) & \scriptsize $\rm{N_{lines}}$\\
    \tiny Sun & \scriptsize Mg I & \scriptsize 7.53 $\pm$ 0.03 & \scriptsize 1 \\
    \tiny Sun & \scriptsize Ca I & \scriptsize 6.36 $\pm$ 0.06 & \scriptsize 7 \\
    \end{tabular}
\tablefoot{Example of two rows from the tables with atomic and final chemical abundances. Top: Atomic lines used for deriving the abundance ratios presented in this work. Their logarithm values of the oscillator strength ($\mathrm{log}\,gf$) and excitation potentials ($E_{low}$) are sourced from the atomic database presented in the study by \cite{2021Heiter}. Middle: Final results from the averaged line-by-line abundances, along with the corresponding uncertainties and the number of lines used for each element. Bottom: Derived absolute solar abundances used in this work to calculate the final bracket abundance ratios, $\left[\mathrm{\frac{X}{Y}}\right]$ 
abundance ratios.}    
\end{table}

We derived the chemical abundances for the following families of elements: $\alpha$ capture (Mg, Si, Ca, Ti); odd-Z (Na, Sc, V); iron-peak (Cr, Fe, Co); $n$-capture (slow-process: Ba, La; rapid-process: Eu). The atmospheric parameters presented in Sect.~\ref{sect:AP} ($\mathrm{T_{eff}}$, $\log\,g$) were fixed to measure line-by-line chemical abundances. To obtain the various chemical abundances we followed the strategy of Sect.~\ref{sect:AP}, i.e., using the Turbospectrum code, the MARCS model atmosphere, and the GES line-list. During the synthesis of line-by-line abundances, we adjusted the line fitting to the different resolutions of the three setups and accounted for broadening effects by fitting the$\mathrm{v_{mic}}$ for each star and determining the $R$ parameters individually for each spectrum, allowing for a local re-normalization. Although the $R$ values for the three GIRAFFE filters are known, we allowed this parameter to vary because, in the \texttt{iSpec} fitting procedure, the $R$ optimizes the combined effects of the star's rotational velocity and macroturbulence \citep{2014A&A...566A..98B}, while accounting for the fact that the three setups possess different resolving powers (listed in Table \ref{tab:obser}).

The line-by-line abundance measurements were averaged for each element and then expressed relative to solar values. For each setup, the solar values of each element were derived from a solar spectrum obtained from the public library of \cite{2014A&A...566A..98B}, which was cut and degraded to match the wavelength range and resolutions of the three different setups. The derived solar abundances are published at the CDS. By adding the absolute solar $\left[\mathrm{\frac{X}{Y}}\right]_{\odot}$ ratios to our element ratios, we derived the final $\left[\mathrm{\frac{X}{Y}}\right]$ abundance measurements with improved internal accuracy.

The combination of the three GIRAFFE setups allows us to measure different chemical species across three wavelength windows of approximately $\sim 240-320$ \AA.  Due to the limited wavelength coverage and varying resolutions, the line selection was customized independently for each setup. Initially, we relied on the quality flags \textit{synflag} (indicating blending properties) and \textit{gf\_flag} (indicating the accuracy of the $\mathrm{log} gf$ value) associated to the GES line-list \citep{2021Heiter}. However, due to the low S/N of some of the targets, we performed a subsequent visual inspection to discard lines with problematic fits, ultimately retaining only one or a few lines per element. The only exception was for Fe I lines, where we removed those across the three setups that showed systematically different results (see sections below for further details). Furthermore, due to the quality of our spectra, some elements could only be measured for a sub-sample of our FLAMES dataset. This restrictive strategy ensures that we obtain trustworthy results for studying the chemical evolution of Sgr.  

The spectroscopic analysis was done assuming local thermodynamic equilibrium (LTE). Hyperfine structure (HFS) and isotopic splitting were taken into account in the analysis using the atomic data compiled from \cite{2021Heiter} and the references therein.

Concerning the uncertainties in the abundances, iSpec employs two internally consistent and homogeneous methods \citep{2014Blanco}. When sufficient spectral lines are available, uncertainties are calculated using the standard error of the mean of the individual line-by-line measurements. If there are not enough lines, the uncertainties were instead determined from the covariance matrix generated by the non-linear least-squares fitting algorithm. This method accounts for how the synthetic spectrum changes when the best-fit parameters are adjusted, as well as the residuals between the synthetic and observed spectra. It is important to emphasize that for many elements (e.g., Mg {\scriptsize I}, Cu {\scriptsize I}, Co {\scriptsize I}, Cr {\scriptsize I}, La {\scriptsize II}, Eu {\scriptsize II}), we were able to find only a few lines, or in some cases, just one line, suitable for abundance determination. The derivation of abundances for these species, which have fewer detected lines, can be more challenging if the only measurable lines are weak, exhibit blends, or are affected by uncertainties in atomic or molecular data. However, the internal uncertainties are not necessarily larger compared to abundances measured from multiple lines (e.g., $\rm{N_{lines}} \gtrsim 3$). In fact, if the selected lines originate from different setups (with varying $R$ and S/N), this can increase the internal abundance spread, affecting the precision of the measurement.

The line selection used in the analysis to determine the final average abundances for the entire sample is available at the CDS, with the first lines presented as an example in Table \ref{tab:lines}.

\subsubsection{Stellar metallicities} \label{sect:metallicities}

The iron abundances were derived for the entire stellar sample, with values spanning from $-2.13$ to $-0.35$~dex. The typical metallicity uncertainties are approximately 0.1~dex. The final analysis employed eighteen Fe lines, sourced from the blue setup HR8 and the intermediate seutp HR11. Lines from HR15 were excluded after it was found they exhibited an offset compared to the other two setups, likely due to their lower spectral resolution (see Sect.~\ref{sec:append_met}).

\begin{figure}[t]
\centering
 \includegraphics[width=\columnwidth]{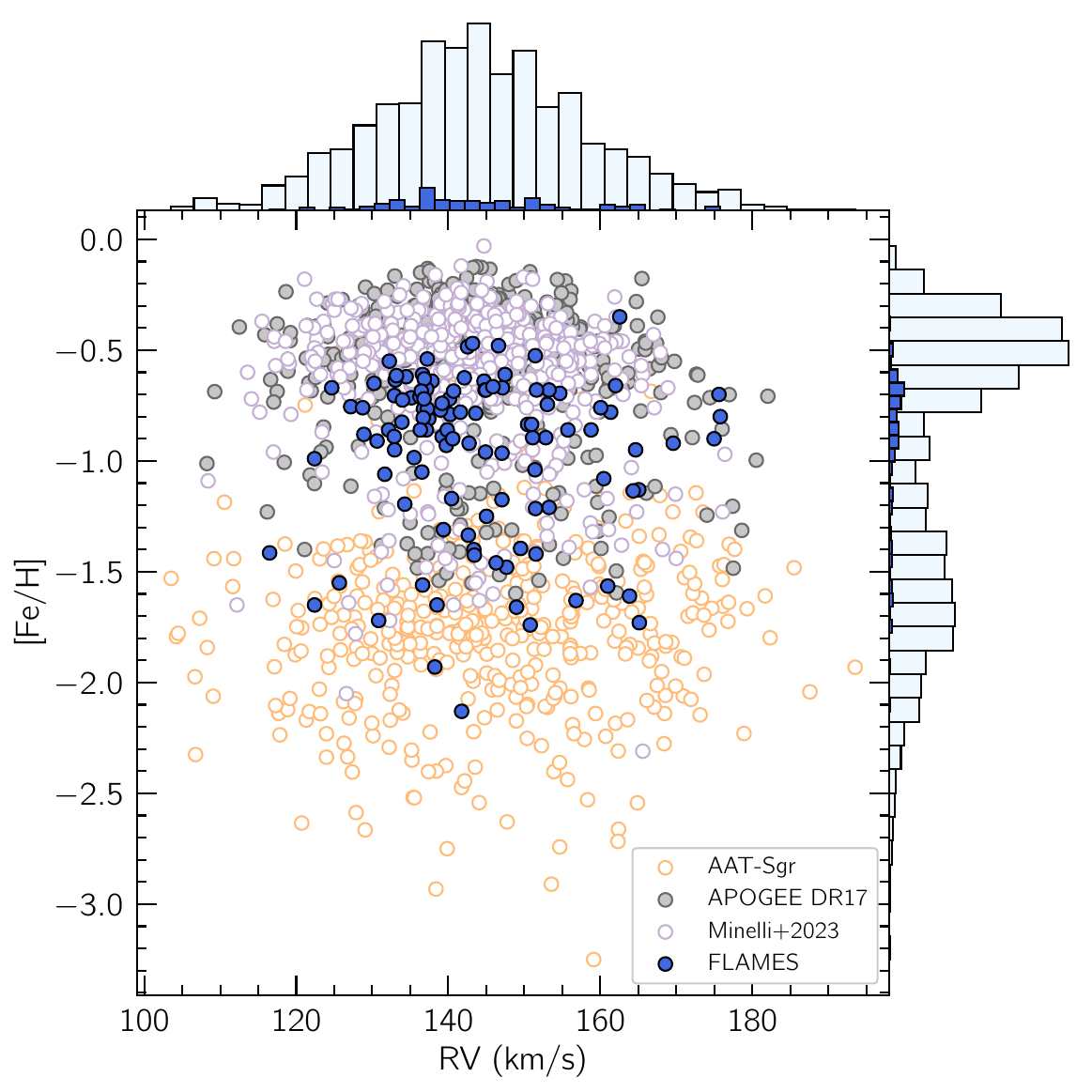} 
\caption{\small{Metallicities and RVs for the Sgr members observed in this study are compared with those from three literature samples. These are the same samples shown in Fig.~\ref{fig:RVhisto}, and the selection criteria applied are similar to those used for the FLAMES targets. Top: Histogram of RVs, with literature data shown in light blue and our targets in blue. Bottom right: Histogram of metallicities, using the same color scheme as the RV histogram.}}
 \label{fig:rv_feh}
\end{figure}

The metallicity distribution, plotted against RV values, is shown in Fig.~\ref{fig:rv_feh} and compared with the same literature samples already reported in Fig.~\ref{fig:RVhisto}. The literature [Fe/H] distribution is bimodal, reflecting the selection biases of various surveys and samples. Our targets are positioned at the tail of the more MR peak, with the majority of stars having [Fe/H] $\sim -0.7$~dex. Although our sample includes fewer stars in the more MP range (77 stars with [Fe/H] $\geq -1.0$~dex), it contains 34 stars with [Fe/H] $< -1.0$~dex among which 13 with [Fe/H] $< -1.5$~dex.

\subsubsection{$\alpha$ elements: Mg, Si, Ca, and Ti}
The $\alpha$ elements are synthesized in the cores of massive stars during the late stages of evolution. Specifically, they are produced and ejected to the ISM in core-collapse supernovae (CCSNe) via different processes \citep{2002Heger,2006Kobayashi, 2013Nomoto}. These processes are expected to enrich the ISM on short timescale \citep{2016Matteucci}. 

Among this group, we were able to measure four different species. Magnesium (Mg) was derived for 82 stars, relying on a single line at 571.10 nm (HR11) that is not affected by blending, resulting in an average abundance of $\left<\mathrm{[Mg\,\text{\scriptsize I}/Fe]}\right> = 0.12 \pm 0.09$ dex. Silicon (Si) measurements were more challenging to derive due to the low S/N of the spectral lines. Final Si I abundances are available for only 17 stars, inferred from two weak lines at 569.04 nm and 579.30 nm (HR11). An attempt was made to use the line at 672.18 nm in the redder setup, but it turned out to be too weak for reliable fitting. The average abundance ratio is given by $\left<\rm{[Si\,\text{\scriptsize I}/Fe]}\right> = 0.18 \pm0.06$ dex. For calcium (Ca) more absorption lines were available throughout the HR15 setup and was derived for 95 targets. The $\left<\rm{[Ca\,\text{\scriptsize I}/Fe]}\right> = 0.17 \pm0.11$ dex is based on six Ca I lines from $\approx 646-671$ nm. Some of these lines are slightly affected by blending effects, which could be contributing to the higher uncertainties. Finally, titanium (Ti) was measured for two different ionization stages. For Ti I, abundances were computed for 78 stars using five lines from three different setups, resulting in an average value of $\left<\rm{[Ti\,\text{\scriptsize I}/Fe]}\right> = 0.11 \pm0.11$ dex. We were able to fit the ionized Ti II line at 649.15 nm, but we discarded it due to poor flags in the GES line list.

\subsubsection{Fe-peak elements: Cr, Co}

Due to the broad range of atomic numbers covered by iron-peak elements (21 $\leq Z \leq 30$), various complex nucleosynthesis processes have been identified as contributing to their production. In our analysis, the Fe-peak elements examined are those elements close to iron in the periodic table, with their primary contribution coming from Type Ia supernovae (SN~Ia). These explosions typically occur in binary systems with a white dwarf as a companion \citep{1995Woosley,2013Seitenzahl} and synthesize these elements on a longer timescale compared to the $\alpha$ elements \citep{2009Matteucci}.

For chromium (Cr), the final results are based on the spectral line at 493.63 nm in HR8 with good GES flags are $\left<\rm{[Cr\,\text{\scriptsize I}/Fe]}\right> = 0.03 \pm 0.11$ dex and are averaged on 73 Sgr stars. Cobalt (Co) abundances were derived for 83 stars using two lines at 677.04 nm and 564.72 nm, both accounting for HFS splitting \citep{1996Pickering}. The resulting average abundance is $\left<\rm{[Co\,\text{\scriptsize I}/Fe]}\right> = 0.02 \pm 0.06$ dex. 

\subsubsection{Odd-Z elements: Na, Al, Sc, V}

Among the odd-Z group, the light elements  sodium (Na) originates in massive stars via hydrogen burning or carbon burning \citep{2015Cristallo,2020Kobayashi}. Aluminum (Al) is noted to be an element whose yields are metallicity-dependent \citep{2013Nomoto}, and at the same time, it is mostly produced in CCSNe at rates comparable to $\alpha$-capture elements \citep{2020Kobayashi}. Heavier elements from this group, specifically scandium (Sc) and vanadium (V), were analyzed in this study. Sc shares similar origins with Na, as it is synthesized during the carbon and neon burning phases in massive stars, while V is primarily produced through silicon burning in Type II supernovae \citep[SNe II,][]{1995Woosley,2015Battistini}.

For the light element Na, we relied on the HR11 setup. We identified two Na I lines at 568.26 nm and 568.82 nm and measured sodium abundances for sixty targets. However, both lines are affected by blending features, and for roughly half of the sample, we were unable to achieve a satisfactory fit. This may be reflected in the overall average uncertainty of the final abundance measurements $\left<\rm{[Na\,\text{\scriptsize I}/Fe]}\right> = -0.18 \pm0.13$ dex. Concerning Al, its derivation posed challenges as we could only use for most of the stars a single absorption line in the redder part of the spectrum (HR15) at 669.86 nm. For a minority of stars, we were also able to measure abundances from a line at 669.60 nm. According to the GES flag only the redder Al line (669.86 nm) is unblended. However, this line was quite weak in our spectra and could only be fitted for a small minority of the sample, specifically 17 stars. Their average abundance was found to be $\left<\rm{[Al\,\text{\scriptsize I}/Fe]}\right> = 0.06 \pm0.06$ dex.

Moving to the heavier elements in this group, Sc was inferred for 62 stars relying on four lines from the HR11 (between 565 and 566 nm) and HR15 (660 nm) setups, resulting in $\left<\rm{[Sc\,\text{\scriptsize II}/Fe]}\right> = -0.04 \pm0.06$. For V, six lines in the central part of the wavelength range (562–573 nm) were fitted for 75 stars providing the mean abundance ratio is $\left<\rm{[V\,\text{\scriptsize I}/Fe]}\right> = 0.06 \pm0.1$ dex. The HFS data were taken from \cite{childs1979laser} and \cite{1998Cochrane}.

\subsubsection{N-capture elements: Ba, La, Eu}

Neutron ($n$)-capture elements are those heavier than zinc ($Z > 30$), formed through the capture of neutrons by a seed nucleus, which subsequently undergoes $\beta$-decays \citep{1957Burbidge, 2008Sneden}. Based on the density of the neutron flux, which determines the rate of nucleosynthesis, these processes are classified as slow ($s$), intermediate ($i$), and rapid ($r$) processes \citep{1977Cowan,1999Busso,2007Kratz}. The $s$-process is known to take place in low- and intermediate-mass stars during their asymptotic giant branch (AGB) phase in longer timescales (see for instance \citealt{2014Bisterzo}) differently to $r$-process that is believed to occur in CCSNe and neutron-star mergers (e.g., \citealt{2018Cote}).

From the $s$-process elements, we derived barium (Ba) abundances using the redder line at 649.68~nm (HR15), which is generally strong in stellar spectra, except in VMP stars, but with reliable $\log gf$ data. However, due to the low resolution of this spectral window for the majority of the targets, we were only able to fit the line correctly for 20 stars, resulting in an average abundance of $\left<\rm{[Ba\,\text{\scriptsize II}/Fe]}\right> = 0.39 \pm 0.06$~dex. We then measured the heavy $s$-process element lanthanum (La) in 73 stars, more than double the number of stars compared to Ba. For these lines, corrections for HFS and isotopic splitting are the ones measured by \cite{1993Villemoes}. The average abundance $\left<\rm{[La,\text{\scriptsize II}/Fe]}\right> = 0.67 \pm 0.10$~dex was computed using a line in the HR8 region at 492.17~nm, which is only slightly affected by blending. As with the Ba line, HFS corrections were necessary in the spectral analysis \citep{Lawler2001}. The only $r$-process element derived for our sample is europium (Eu). Although Eu abundances are typically measured from the bluer part of stellar spectra ($\lessapprox420$ nm), the Eu {\scriptsize II} line at 664.51~nm in the HR15 region can be measured in giant stars with higher S/N. Despite the inclusion of HFS \citep{1993Villemoes} and isotopic corrections \citep{Lawler2001}, which reduce the impact of blending effects, Eu measurements remained difficult and were only possible for 29 stars, with an average abundance of $\left<\rm{[Eu,\text{\scriptsize II}/Fe]}\right> = 0.40 \pm 0.07$~dex.

We attempted to measure some of the light elements belonging to the $s$-process family, such as yttrium II at 508.74~nm and zirconium {\scriptsize I} and {\scriptsize II} at 507.82~nm and 511.22~nm, respectively. However, no successful spectral synthesis yielded satisfactory results.

\section{Results and discussion} \label{sect:discussion}

\subsection{Comparison with other Sgr samples}\label{sect:comparison_sgr}

\begin{figure*}[!t]
\centering
 \includegraphics[width=\textwidth]{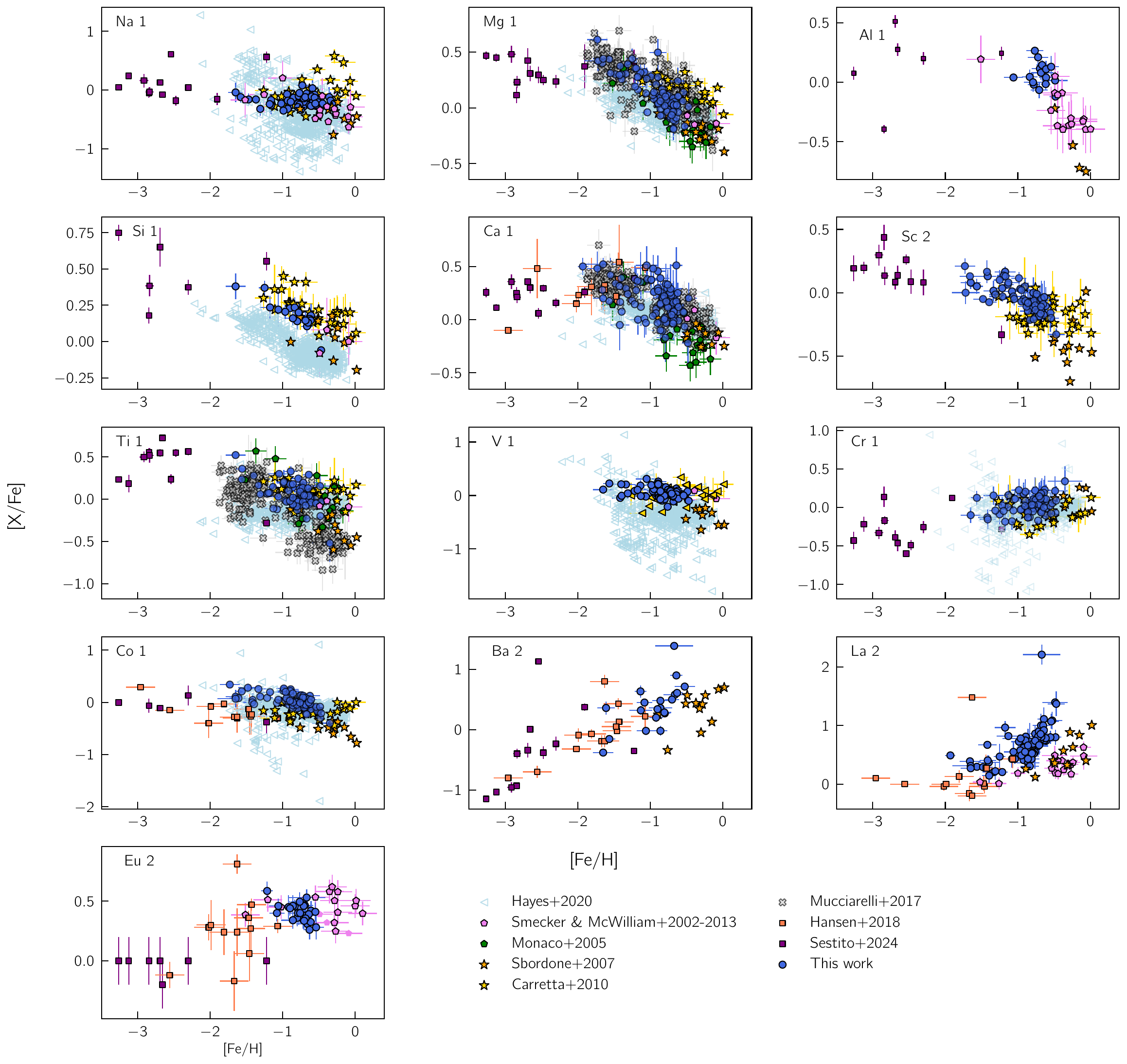} 
\caption{\small{Chemical abundances [X/Fe] as a function of [Fe/H] for different Sgr samples. The FLAMES targets analyzed in this work are represented as blue circles. Different literature samples, represented by distinct colors and markers, provide additional [X/Fe] data. These include the APOGEE dataset for the Sgr main body from \cite{2020Hayes}, and optical data for the Sgr core from \cite{2005Monaco}, \cite{2007Sbordone}, \cite{2010Carretta}, and \cite{2017Mucciarelli}, with the most metal-poor stars reported by \cite{2018Hansen} and \cite{2024Sestito}. The studies by \cite{2002Smecker} and \cite{2013McWilliam} contain data for both the nuclear cluster M54 and the main body of the galaxy.}}
 \label{fig:chemical_trend}
\end{figure*}

Figure~\ref{fig:chemical_trend} displays our PIGS-HR-Sgr element abundances (blue points) compared with the Sgr core selection performed by \cite{2020Hayes} on the near-infrared (NIR) APOGEE data (light blue triangles). The comparison is further extended to include five additional high-resolution optical spectroscopic samples and one medium-resolution sample, all focusing on the Sgr core. 

The studies by \cite{2005Monaco} and  \cite{2007Sbordone} conducted with FLAMES-UVES and UVES respectively, include 15 and 12 RGB Sgr stars with $-1.0 \lesssim \rm{[Fe/H]} \lesssim 0.0$, with one star at [Fe/H]~$= -1.52$~dex \citep{2005Monaco}. Both catalogs include abundances for $\alpha$, Fe-peak and heavy elements. Again with FLAMES, \cite{2010Carretta} inferred the abundances of various elements in 27 giant stars ranging from solar metallicity to $-1.2$~dex in the core of the Sagittarius galaxy. The mid-resolution sample of \cite{2017Mucciarelli} comprises 235 Sgr giant stars (light gray crosses) from both the main body and the nuclear cluster M54. Iron and $\alpha$ element abundances are provided across a wide metallicity range ($-1.8 \lesssim \rm{[Fe/H]} \lesssim -0.1$). Lower metallicity regimes are covered by the studies of \cite{2018Hansen} and \cite{2024Sestito}. The former includes 13 different chemical species for 13 giant stars with $-2.97 < \rm{[Fe/H]} < -1.07$~dex, while the latter analyzed 12 low-metallicity Sgr stars with [Fe/H] down to $-3.26$~dex, providing a detailed chemical analysis. Lastly, we included the studies by \cite{2002Smecker} and \cite{2013McWilliam}. From \cite{2002Smecker}, we obtained Na, Al, La and Eu measurements for 14 RGB stars in Sgr, while from the \cite{2013McWilliam}, we retrieved 3 RGB stars in the M54 cluster for the remaining elements (pink pentagons in Fig.~\ref{fig:chemical_trend}). We emphasize that the literature samples introduced in this section, primarily covering the core of Sgr, do not directly cross-match with our targets, as only a small number of stars are common to both datasets. In fact, we found no overlapping targets with the optical literature samples introduced above. Only seven stars are in common between our PIGS-HR-Sgr sample and the one from \citet{2020Hayes}. However, a one-to-one comparison is uncertain, as all these stars are flagged in terms of their parameters and S/N in APOGEE DR17. Therefore, we use the comparison with literature samples as a general assessment of the overall trends and distributions for Sgr, which can help to further interpret our results below.

Examining the different panels of Fig.~\ref{fig:chemical_trend}, our data align with most literature trends. The $\alpha$ elements exhibit the expected behavior with respect to metallicity. In fact, both our results and those from the literature show the typical decline in $\alpha$ abundances as metallicity increases, which is attributed to the time delay between prompt SNe II and SNe Ia, acting on longer timescales. The most noticeable differences are the ones with the APOGEE data, particularly for elements such as Si {\scriptsize I}, and less for Ti {\scriptsize I}. The NIR wavelength coverage of APOGEE (as opposed to the optical range used for the other samples) partly explains these discrepancies, as the offset in NIR Ti {\scriptsize I} abundances may be influenced by problematic lines or their dependence on uncalibrated $\rm{T_{eff}}$ values \citep{2020Jonsson}. This discrepancy with APOGEE, which becomes larger for the most iron depleted stars, is also observed in the samples from \cite{2005Monaco, 2010Carretta} and \cite{2017Mucciarelli}, while the Ti abundances reported by \cite{2007Sbordone} are lower and more comparable to those of APOGEE, though they cover the most MR end of the distribution. The results of \cite{2024Sestito} account for NLTE corrections which range from 0.1 ([Fe/H] $\lesssim -0.5$~dex) to 0.3~dex for the most MP targets ([Fe/H] $\lesssim -1.5$~dex) according to \cite{2011Bergemann}. The other samples do not.

Si also shows an offset with the APOGEE measurements, which is expected to be very precise \citep{2023Heged}. We therefore believe that our measurements, based on two weak absorption lines, may exhibit a systematic offset compared to the APOGEE measurements. This offset ($\sim 0.24$ dex) could be due to differences between optical and infrared observations or line-list systematics. We consulted the MPIA database\footnote{\url{https://nlte.mpia.de/}} to gather NLTE corrections for the Si {\scriptsize I} lines used \citep{2013Bergemann}. However, the maximum correction of 0.03 dex is insufficient to account for the offset observed in our results. On the other hand, our trend aligns much more with the optical samples from \cite{2007Sbordone} and \cite{2010Carretta} and all show the typical decline of $\alpha$ elements with metallicity.

Mg and Ca show the most prominent drop with metallicity (up to $\sim 1$~dex) and the trends for our FLAMES targets are well aligned with those of the other optical Sgr samples but deviate more from the Hayes sample. Since APOGEE applies NLTE corrections for Mg and Ca \citep{2020Osorio, 2020Jonsson}, we verified that applying corrections from the MPIA database \citep{2015Bergemann} of $\sim-0.2$~dex for the Mg line at 571.10~nm would align our results with the APOGEE measurements. Regarding Ca, we were unable to find NLTE corrections for the lines used in this study.

Analyzing the overall trend of Na from the FLAMES sample along with three other optical samples, it appears that Sgr shows a decline in [Na/Fe] with increasing metallicity. This trend is expected, as Na is predominantly produced by massive stars \citep{1995Woosley}. Our values do not display such strong decline nevertheless they are comparable to the compilation reported by \citeauthor{2013McWilliam} \citeyear{2013McWilliam} (see their Fig.~ 4). The measurements from \cite{2010Carretta} exhibit a different behavior, suggesting an increase for more metal-rich (MR) stars, making the overall trend harder to interpret. NLTE corrections that could be applied to our lines might lead to adjustments of $-0.05$ to $-0.1$ dex \citep{2022Lind}, which would bring our results closer to those of \cite{2007Sbordone}. The results for Al, derived using LTE for all samples, are presented in the first row. Despite a slight spread at lower metallicity, a clear, tight trend highlights the primary origin of this element, predominantly from CCSNe \citep{2020Kobayashi}. However, the large variation of nearly 1 dex in [Al/Fe] could be attributed to the strong metallicity dependence of this element \citep{2013Nomoto}. Nevertheless, a clear trend remains, reflecting the primary source of this element, which is mostly CCSNe \citep{2020Kobayashi}. AGB stars may also contribute to Al production, which aligns with the previously studied scenario that these stars played a role in the chemical enrichment of Sagittarius \citep{2017Hasselquist,2018Hansen, 2024Sestito}.

The heavier odd-Z elements, Sc and V, follow different trends with metallicity. Sc shows a clear decrease in abundance with increasing [Fe/H], a trend consistent with the findings of \cite{2007Sbordone} and \cite{2010Carretta}. This behavior aligns with Sc’s primary production by CCSNe. In contrast, V exhibits a less pronounced decrease, though there is a general downward trend when considering both the FLAMES sample and the work of \cite{2007Sbordone}. For this element, APOGEE is known to have low reliability in its abundance measurements \citep{2021Buder, 2023Heged}.

The two Fe-peak elements align well with the literature. The increase in Cr with metallicity matches expectations, as its primary source is SN~Ia \citep{2020Kobayashi}. In contrast, Co shows a more variable trend due to its more inhomogeneous origins, which include both CCSNe and, to a lesser extent, AGBs \citep{2013Nomoto,2020Kobayashi}. Although NLTE corrections can affect the abundances of Cr and Co lines by 0.1 to 0.2 dex \citep{2010Bergemannces, 2010BergemannCo}, we did not apply these corrections because the comparison samples also assume LTE in their analyses. 

The three last elements presented in Fig~\ref{fig:chemical_trend} belong to the $n$-capture elements. As previously mentioned, the heavy $s$-process elements (or second-peak elements), being released over longer timescale by AGB stars, exhibit an increase over time and with metallicity. This trend is evident for both Ba and La and is consistent across all samples reported in the figures, collectively showing an increase of nearly 2~dex for both elements. It is important to note that for Ba, we used the line at 649.68~nm, which is known to be affected by NLTE effects (see, e.g., \citealt{2015korotin}). These effects can cause a variation in A(Ba) of about $-0.2$~dex for our atmospheric parameter space \citep{2019Eitner}. However, this variation does not eliminate the overall increasing trend with [Fe/H]. Since only \cite{2024Sestito} accounts for these corrections, we retain the abundances on the LTE scale for consistency with other samples. The PIGS-HR-Sgr La measurements, which cover a broad range of metallicities, show a sharper increase compared to the results of \cite{2018Hansen}, which only begin to rise after leveling off for [Fe/H]~$< -2.0$~dex. In addition, the samples from \cite{2007Sbordone} and \cite{2002Smecker} display a trend very similar to ours.

One star in our sample stands out with significantly higher La and Ba abundances, enhanced by $\approx$ 1–1.5 dex relative to the sample average. This star shows no distinct characteristics from the rest of the Sgr sample in the parameter or other chemical spaces. To confirm its membership in the Sgr core, we examined its RV and proper motion values, finding them to validate the Sgr origin of the target. These checks, combined with a visual inspection of the best fits to the La {\scriptsize II} and Ba {\scriptsize II} abundances, confirmed the genuine enhancement of heavier elements in this star, though this enhancement does not extend to the $r$-process element Eu. Generally, these overabundances of elements synthesized by the slow $n$-capture process in stars such as giants, which cannot produce $s$-process elements internally at their evolutionary stage, are thought to result from binary interactions \citep{1957Burbidge, 2011Kappeler}. We examined the RV variability of this target by comparing the RV values derived from three different GIRAFFE setups and comparing these results to a star with "normal" heavy element abundances. No RV variability was detected beyond the precision limits of GIRAFFE, and the RV curves for both the potential binary target and the comparison star were consistent. However, this is not sufficient to exclude the possibility of a binary nature for the star, as the orbital period could be very long, potentially reaching up to 30,000 days \citep[e.g.,][]{2019Escorza, 2019Jorissen} and thus not detected by our few RV measurements. To further investigate the potential binarity of this Sgr giant and achieve a more accurate and complete chemical characterization, a broader wavelength coverage would be required.

The spread in europium (Eu) abundances highlights the challenges in accurately measuring this element. Although the data from \cite{2024Sestito} do not reveal a clear trend with [Fe/H], when considering the entire optical sample, there is an observable increase of approximately 0.5~dex. If we exclude the literature measurements from \cite{2024Sestito} and \cite{2018Hansen}, which cover the most metal-poor regime, our FLAMES data, along with those of \cite{2002Smecker} (pink pentagons), level off around [Eu/Fe] $\sim 0.5$ dex. This rise in Eu abundances is probably an indication of a prolonged and later active star-forming period, during which massive stars may have continued enriching the ISM with $r$-process elements \citep{2021Reichert}. Other exotic scenarios have also been proposed to explain the increase in $r$-process elements at higher metallicities, such as neutron star mergers (NSMs) contributing with a delayed production of Eu \citep{2018Cote,2020Skuldelayed} or massive stars that formed and exploded as magneto-rotational supernovae \citep[MR-SNe,][]{2021Reichert}. This scenario will be further investigated in Sect.~\ref{sect:heavy}. 

\subsection{$\alpha$-knee}\label{sec:knee}

\begin{figure}[t]
\centering
 \includegraphics[width=\columnwidth]{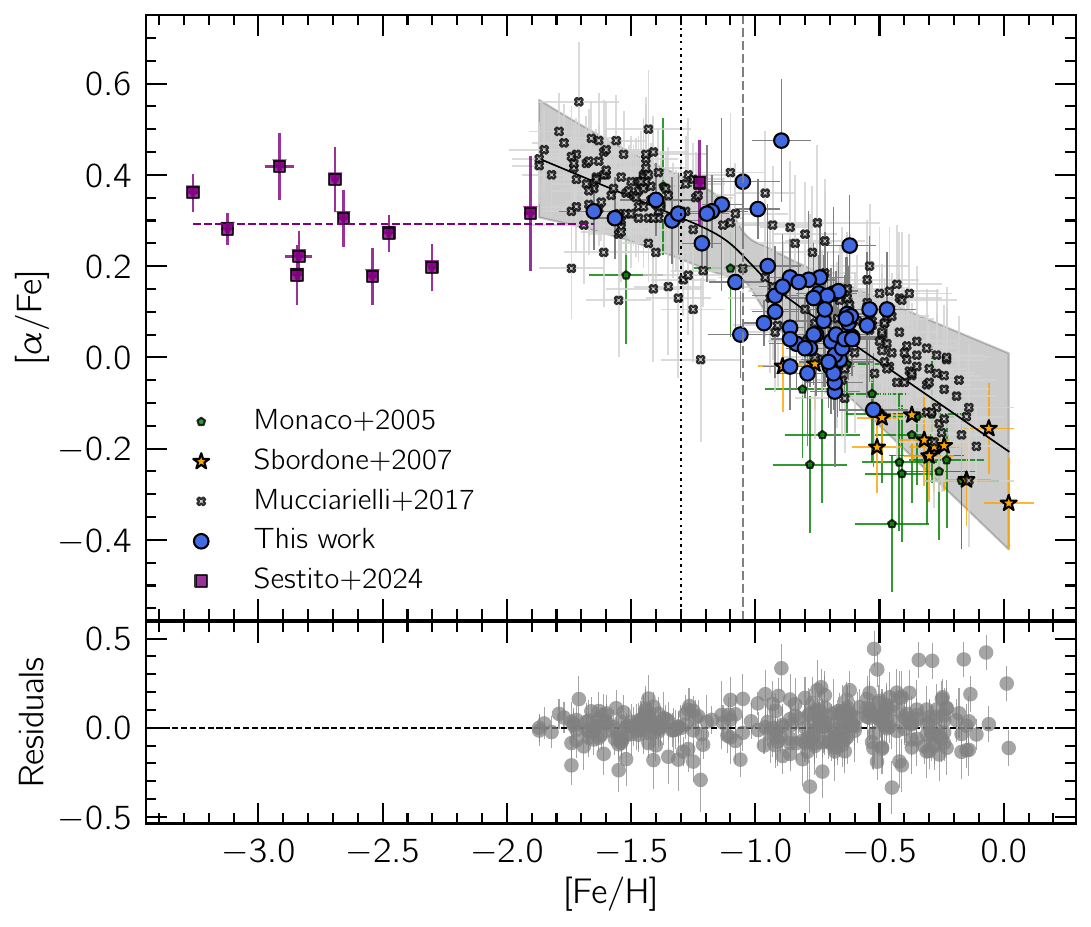} 
\caption{\small{Trend of the $\alpha$ elements vs. [Fe/H] for the FLAMES targets (blue circles), compared with literature samples from \cite{2005Monaco}, \cite{2007Sbordone}, \cite{2017Mucciarelli}, and \cite{2024Sestitocar}. For all the samples, Mg and Ca abundances, propagating their corresponding uncertainties, were averaged and then fitted with two linear functions connected by a smoothing function. The gray shading represents the fit uncertainties, which are also displayed as residuals in the bottom plot. As explained in the text, the data from \cite{2024Sestitocar} were excluded from the fitting procedure. The dashed gray line marks our results for the knee position at [Fe/H]~$= -1.05$, while the dotted black line at [Fe/H]~$= -1.3$ represents an approximation of the results from \cite{2010Carretta} and \cite{2014deBoer}}} 
 \label{fig:knee}
\end{figure}

In Fig.~\ref{fig:knee}, we further compare the well-known $\alpha$-knee diagram, or Tinsley diagram \citep{1979Tinsley}, which is often used as an effective tracer of early star formation efficiency in galaxies. The dominance of the yields of CCSNe during the early chemical enrichment results in the production of $\alpha-$elements. This is followed, after a time delay, by the metal production from SN~Ia, which predominantly generates iron-peak elements. Therefore, the metallicity at which the turnover in the [$\alpha$/Fe] - [Fe/H] occurs is linked to the star formation rate of a galaxy. A lower metallicity suggests a galaxy with a less efficient SFH as fewer SNe~II events balanced the delayed contribution from SN~Ia. In contrast, galaxies that build up and retain metals and gas quickly before the contribution of SN~Ia to chemical enrichment will reach a higher metallicity at the $\alpha$-knee (e.g., \citealt{2006Kobayashi, 2020Kirby}). 

For our data the fitting was performed using the \texttt{scipy Python} function \texttt{curve\_fit}. To increase the statistical significance, we complemented our sample with three literature samples from \cite{2005Monaco, 2007Sbordone,2017Mucciarelli} covering similar metallicity range. As in the comparison studies, we use an average of Mg and Ca as a proxy for $\alpha$ abundances, considering only the stars for which both Mg and Ca could be measured. Namely, we used a weighted average, as it provides a more reliable proxy for $\alpha$ abundances than relying on a single Mg line \citep[as discussed in][]{2017ASInC..14...37J}. We applied a regression model consisting of two linear functions connected by a sigmoid smoothing function to fit the knee point at $\rm{[Fe/H]_{\alpha-knee}}$ = $-1.05\pm 0.13$. The residuals of the fit are shown in the bottom panel of Fig.~\ref{fig:knee}. The data from \cite{2024Sestito} are displayed on the same figure, covering a different metallicity range. Although not statistically sufficient for fitting, this dataset serves as a visual guide to the average trend for metallicities below [Fe/H]~$= -1.0$. If SN~Ia were already influencing at this metallicity regime, we would expect a distinct slope, which is hinted at by the data from \cite{2017Mucciarelli}. In the more metal-poor range observed by \cite{2024Sestito}, the data appear to level off, suggesting a low contribution of SN~Ia to the chemical enrichment in this metallicity regime. Moreover, when examining our FLAMES data (shown in blue in Fig.\ref{fig:knee}), we can distinguish between the main stellar population, characterized by [Fe/H] $\sim -0.8$, and the remaining older population with [Fe/H] $< -1.0$. This metal-poor population shows a plateau at $\rm{[\alpha/Fe]} \approx 0.3$, consistent with the trend observed in the more metal-poor stars studied by \cite{2024Sestito}. This behavior suggests the absence of significant Type~Ia SNe contributions in this metallicity regime. Additional data for [Fe/H]$ < -1.0$ and the comparison with chemical evolution models would help clarify the timing of SN~Ia onset in this regime.

It is interesting to note that, in the third panel of the first row of Fig.~\ref{fig:chemical_trend}, the [Al/Fe] values begin to flatten at $\rm{[Fe/H]_{\alpha-knee}} = -1.05$, corroborating the occurrence of the knee around this metallicity. However, the lack of higher statistics makes it impossible to exclude the possibility of a shift toward slightly lower metallicities, as suggested by \cite{2017Mucciarelli}. Additionally, we acknowledge that the scatter in the abundance data makes the shape of the $\alpha$ knee highly sensitive to the smoothing parameter, an effect that is reflected in the error associated with the knee estimation. Also, the paucity of data below [Fe/H] $<-1.0$ and the low number of stars at the location of the knee affect the precise determination of the knee position. Another contributing factor is that the compilation of the dataset presented in Fig.~\ref{fig:knee} is inhomogeneous in terms of resolution and instrumentation, which can introduce systematics into the comparison, as well as differences in spatial sampling. In fact, in our case, we are examining a very spatially confined dataset located in the very central region of the core of Sgr. 

\begin{figure}[!t]
\centering
 \includegraphics[width=\columnwidth]{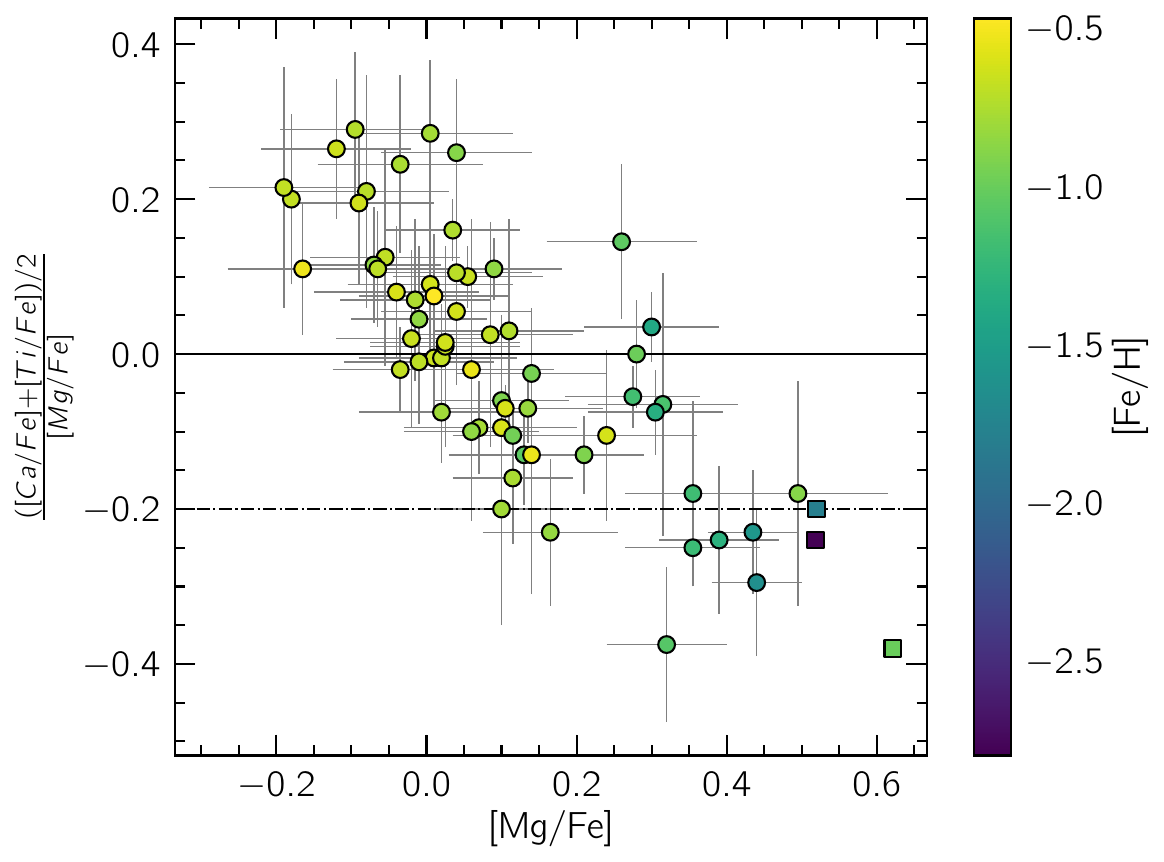} 
\caption{\small{Ratio of the average of the explosive $\alpha$ elements (Ca and Ti) to the hydrostatic $\alpha$ elements (Mg) as a function of [Mg/Fe]. The color code encodes the metallicity of the Sgr stars (filled circles). The square symbols denote the theoretical models from \cite{2013Nomoto}, which incorporate yields from both SNe II and HNe.The dashed-dotted line delineates the region of chemical space where stars could have originated from gas enriched by both CCSNe and HNe.}}
 \label{fig:delta_alpha}
\end{figure}

An additional factor that influences the determination of the knee is the inclusion of NLTE corrections. In the results presented in the main text, we used our LTE measurements, as the comparison samples, except those from \citet{2017Mucciarelli} and \citet{2024Sestito} (which are not included in the fitting procedure), do not account for NLTE corrections. However, we assess the impact of NLTE effects on the knee estimation and present the corresponding results in  Appendix~\ref{sec:knee_nlte}.

Despite these limitations, our result lines up with the finding of \cite{2014deBoer}, who determined the $\alpha$-knee at $\rm{[Fe/H]_{\alpha-knee}} = -1.27$ for the stellar streams of Sgr. We also compared our result with that obtained using the formula presented by \cite{2020Reichert} (equation 6), which posits a dependence of the position of the $\alpha$-knee on the luminosity (and thus the mass) of the galaxy. By applying this formula, we found comparable results within the uncertainties: $\rm{[Fe/H]_{knee, estimates}}$ = $-1.65\pm 0.48$ (using the absolute visual magnitude $M = -13.50$ from \citeauthor{2012McConnachie} \citeyear{2012McConnachie}). This corroborates the idea that smaller systems with lower initial amount of Fe exhibit the knee position at lower metallicities compared to more massive galaxies. \cite{2010Carretta} confirmed the knee at [Fe/H] $\sim -1.3$ dex, occurring at a lower metallicity than our result, which can be attributed to their use of data from M54 and the averaging of Si, Ca, and Ti to represent $\alpha$ element abundances. \cite{2017Mucciarelli}, averaging Ca and Mg found a much more metal-rich knee at $\rm{[Fe/H]\sim-0.4}$ dex. The extensive work of \cite{2021Hasselquist} compares abundance trends for different MW satellites using chemical evolution models. Among these, they analyzed Sgr and found that $\alpha$ abundances flatten at -1.0 $\lesssim$ [Fe/H] $\lesssim -0.8 $, a result that is slightly more MR compared to our findings. However, according to both our results and theirs, Sgr differs from the LMC and SMC. Despite being $\sim$ 100 times more massive, the knee of the LMC is found at lower [Fe/H] values (-2 $\lesssim \rm{[Fe/H]}\lesssim -1.8$ \citealt{20Nidever,2021Hasselquist,2024Chiti})due to its isolated existence. On the other hand, the literature results for the GES system indicate that $\rm{[Fe/H]_{\alpha-knee} \sim -1.2}$ \citep[e.g.,][]{2019Mackereth,2023Fernandes}, suggesting that Sgr and GES share similar early SFH, characterized by strong star formation efficiencies. The different values reported in the literature highlight the challenges of estimating the location of the knee, including the strong dependence on various factors such as the choice of alpha elements, the level of precision in the abundances, and the metallicity coverage of the sample.

Another insightful diagnostic diagram involving $\alpha$ abundances is shown in Fig.~\ref{fig:delta_alpha}, based on the recent study by \cite{2024Bandyopadhyay}. The main idea is to use the ratio between hydrostatic burning (in our case Mg) and explosive nucleosynthesis (in our case Ca and Ti, we do not include Si as it is computed only for a minor part of the sample) $\alpha$ elements and no Fe dependence. The underlying concept is that the mass of the CCSNe progenitor influences this ratio. Specifically, more massive stars (M $> 35 \rm{M_{\odot}}$) contribute to the abundance of hydrostatic elements like Mg, while less massive stars (M $< 25 \rm{M_{\odot}}$) are responsible for explosive elements like Ca and Ti \citep{1995Woosley, 2020Kobayashi}. This suggests that the observed trends are not solely due to the onset of SN~Ia, which would reduce the Mg abundance in more MR stars, but also reflect the contribution of slightly less massive CCSNe, which could influence these trends \citep{2018McWilliam}. Moreover, \cite{2024Bandyopadhyay} proposed a limit of $\Delta\alpha < -0.2$, suggesting that stars with high abundances of hydrostatic elements could originate from gas enriched by hypernovae (HNe) and are represented in Fig.~\ref{fig:delta_alpha} as filled squares. These yields overlap with the locus of the few FLAMES targets exhibiting high Mg abundances, reinforcing the possibility that some of the metal-poor stars in Sgr were enriched by HNe.

\cite{2018Carlin} investigated this same $\alpha$ ratio for stars located in the Sgr stellar streams. Their analysis revealed a lower abundance of hydrostatic elements compared to explosive elements. By comparing these results with those from the LMC and the Fornax dSph, and considering the observed pattern indicating a dearth of massive stars, they could confirm the top-light IMF hypothesis proposed by \cite{2013McWilliam} and \cite{2017Hasselquist}. Their reported $\Delta\alpha$ values appear to be comparable with ours, which is expected given that the streams and the core are components of the same progenitor. These hypotheses could still accommodate the presence of rare hypernovae, which would primarily affect the more MP (and possibly older) stars \citep{2024Lee}. According to the detailed analysis by \cite{2024AsaSk} on Sculptor, the contribution of HNe becomes significant at much lower metallicities. Similarly, contributions from HNe have also been  detected in the low-metallicity regime of Sagittarius from the pattern of the lighter elements \citep{2024Sestito} and from its average [C/Fe], which is lower in this system than in the MW \citep{2024Sestitocar}.

\subsection{Chemical evolution of the biggest Milky Way satellites}\label{sect:satellites}

\begin{figure*}[!t]
\centering
 \includegraphics[width=\textwidth]{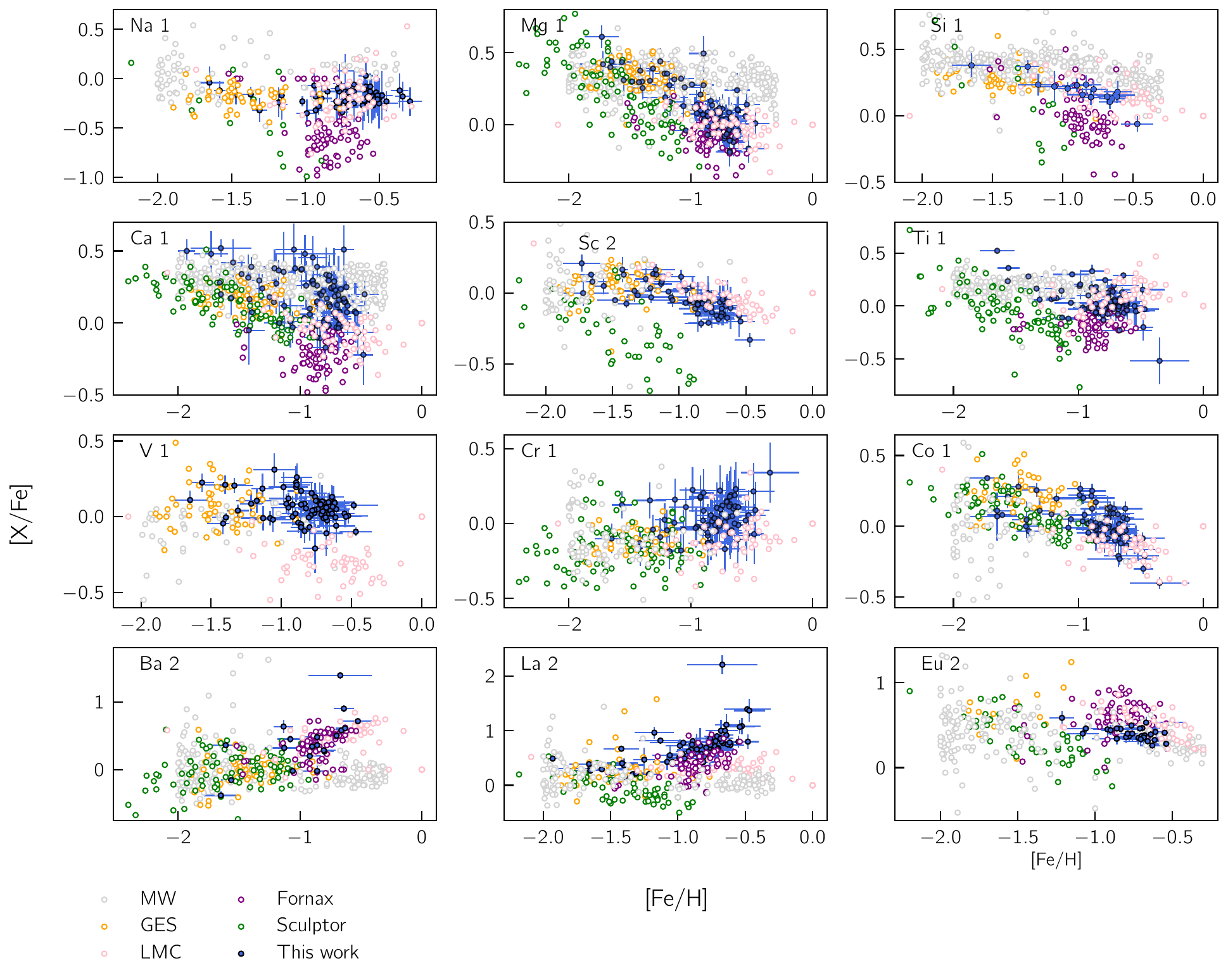} 
\caption{\small{Chemical abundances [X/Fe] as a function of [Fe/H] for the Sgr sample (filled blue dots) are compared with MW halo stars and members of various MW satellites. Specifically, data for the MW halo (empty gray dots) are queried from the SAGA database \citep{2008Suda}. The GES stars (orange) are taken from \cite{2022Carrillo}, LMC members (pink) from \cite{2013Swaelmen}, and Na measurements from \cite{2008Pompeia}. Finally, stars from the Fornax galaxy (purple) are from \cite{2010Letarte}, while Sculptor members (green) are from \cite{2019Hill}.}}
 \label{fig:chemical_trend_mw}
\end{figure*}

To gain deeper insights into the chemical enrichment of Sgr, we examine the same chemical trends shown in Fig.~\ref{fig:chemical_trend} this time comparing them with samples from various MW satellites/structures and MW halo stars. Specifically, for the MW stars we used the Stellar Abundances for Galactic Archaeology database\footnote{Version April 10, 2023.} \url{hhttp://sagadatabase.jp/} \citep[SAGA,][]{2008Suda}. From this database, we queried stars with atmospheric parameters similar to our sample, restricting our selection to giant stars to minimize potential systematic effects in the comparison. To ensure a sufficient number of data points for the comparative analysis, we set the maximum queried resolution to 40,000, even if it is higher than that of GIRAFFE.

We chose to use two massive systems for comparison: 62 GES members from the work of \cite{2022Carrillo}, and 126 LMC stars presented in the studies of \cite{2013Swaelmen} and \cite{2008Pompeia}. Although these systems are among the most massive structures that have interacted or will interact with the MW, their estimated masses and evolutionary histories differ significantly. Indeed, GES (with an estimated total mass of $\sim 10^{9}\,\rm{M_{\odot}}$, e.g., \citealt{2020Mackereth,2020Feuillet}) was fully disrupted following its merger with the MW $\sim 8-11$ Gyr ago \citep{2018Belokurov,2018helminat,2020Das}. In contrast, the irregular and more massive LMC galaxy (total mass of $\sim 10^{10-11}\,\rm{M_{\odot}}$, \citealt{2019Erkal,2021Shipp}) has evolved in almost near isolation and is likely encountering the MW for the first time \citep{2007Besla}.

The other two objects of comparison, Fornax and Sculptor, are dSph galaxies. For the former, the data are taken from \cite{2010Letarte}, while for the latter, they are taken from \cite{2019Hill}. Fornax is the second most luminous and massive dSph galaxy in the Local Group (after Sgr), with an estimated total mass of $\sim 1.6\times10^{8}\,\rm{M_{\odot}}$ \citep{2012DeBoerFor,2009Lokas}, showing a metallicity range comparable to that of Sgr. On the other hand, Sculptor, is less massive ($\sim 3\times10^{7}\,\rm{M_{\odot}}$ \citealt{2009Lokas}) and significantly more metal-poor ($ \langle [\rm{Fe/H}] \rangle \sim -1.8$  \citealt{2019Bettinelli, 2012deboersc}). 

The different [X/Fe] vs [Fe/H] trends are displayed in Fig.~\ref{fig:chemical_trend_mw}. It is important to note that, as with the comparison discussed in Sect.~\ref{sect:comparison_sgr}, there are inherent caveats when confronting results from different studies. These stem from the use of different instruments for observations and diverse analytical approaches (line selection, solar abundances, etc.). Hence, systematic offsets are likely present in these trends. Finally, since no measurements for Al were available from the comparison samples, this element was excluded from the figure. From the various panels, we found that for most elements, Sgr exhibits similarities to the MW halo sample and the more massive systems, GES and LMC. As expected, Sculptor shows the most significant differences, likely due to its mass and the lower metallicities of its stellar populations. Although less pronounced than in Sculptor, Fornax also displays some differences. However, we were not able to obtain measurements for all abundances in this system.

The abundance trends analyzed in this study do not show signs of multiple sequences, indicating that the Sgr stars were formed from the galaxy's own gas, without significant contamination from the MW. This conclusion is consistent with the spatial distribution of the targets, located near the nuclear cluster in a region where the gravitational potential well of Sgr makes it unlikely for external gas to have infallen. Consequently, the chemical enrichment revealed by the analysis reflects the undisturbed chemical history of stellar populations that evolved within the system, following a SFH characterized by multiple star formation bursts, where the most intense ones led to the formation of an intermediate-age population of $\sim8$ Gyr (as will be further outlined in Sect.~\ref{sect:AMR}), consistent with findings from other studies \citep[e.g.,][]{1999Bellazzini,bellazzini2006age,Siegel07}.

Regarding the $\alpha$ elements, as noted in \cite{2022Carrillo}, we find that all the satellites exhibit lower abundances compared to the MW sample, especially for the more chemically evolved stars. This suggests a lower star formation rate or a slower process of chemical evolution and enrichment. The deficiency relative to the MW is particularly evident in the hydrostatic element Mg, which may indicate a lack of more massive SNe II enriching the ISM of the satellites compared to that of the MW, and, in the case of Sgr, it was connected to a top-light IMF by different studies \citep[e.g.,][]{2013McWilliam, 2017Hasselquist}. It is worth emphasizing that differences in the levels of Mg and Ca are challenging to interpret, as they may result from the system’s star formation rate or the yields of SN~Ia \citep{2009Matteucci,2009ApJ...707.1466K,2020Kobayashi}. Consequently, linking the different enrichment levels between Sgr and the MW solely to differences in the IMF is not straightforward.

Among the satellites, there are also differences, with Sculptor and Fornax displaying the lowest $\alpha$ abundances, particularly under-abundances in Mg, Si, and Ca. This difference is less pronounced when examining the GES dataset. According to \cite{2021Hasselquist} this can be explained by the fact that, compared to more isolated dSphs, which exhibit weaker earlier star formation periods, the GES showed enrichment to higher levels before the onset of SN~Ia.

Unlike \cite{2007Sbordone} and \cite{2017Hasselquist}, but similar to \cite{2013McWilliam}, we find that for Sgr, the light odd-Z elements (Na, Sc, V) generally follow similar trends to those of the MW, GES, and LMC, with two exceptions: the V abundance in the LMC and a slight depletion in Na abundances in Sgr compared to the MW. The former difference can be attributed to the HFS corrections adopted by \cite{2013Swaelmen}, which lower the V abundances for their LMC sample by $\sim0.2$ dex, accounting for the discrepancy observed in Fig.~\ref{fig:chemical_trend_mw}. Regarding the slightly higher Na abundances in the MW sample ($\sim 0.2-0.3$ dex), since Na is synthesized during the hydrostatic stages of massive stars in a mass-dependent process, this difference may suggest a lower contribution from higher-mass stars to chemical enrichment in the massive satellites compared to the MW \citep[as found, for instance, in][]{2022MNRAS.510.2407B}, or different mass ranges of the supernovae responsible for Na production. \cite{2004Shetrone} explained the lower Na trends found by compiling Sgr literature data, attributing them to an excess of Fe produced by SN~Ia, which is not expected to produce significant Na. At last, since we did not observe increasing trends with metallicity, we found no evidence of an AGB contribution to the Na abundances. For Sc, compared to Sculptor, the [X/Fe] ratio at the same metallicity is higher in the more massive systems, including Sgr. This suggests a relatively higher contribution of SN~Ia than SNe II in this less massive dSph. 

For the Fe-peak element Cr, the increasing trend with [Fe/H] observed across all samples indicates the significant role of SN~Ia in producing this element. In contrast, Co, belonging to the same category of Cr, shows a much more moderate contribution from this type of supernova, both in all systems and in the MW.

Finally, concerning the neutron-capture, specifically the $s$-process elements Ba and La \citep[which are synthesized in both the $s$- and $r$-processes,][]{2014Bisterzo}, we observe a sharp enhancement ($\sim 0.4-0.5$ dex) in the Sgr abundances compared to those in the MW. This feature has been previously noted and discussed as a clear indication of substantial AGB star contributions \citep{2013McWilliam, 2017Hasselquist, 2021Hasselquist,2024Sestito} with respect to the MW. 

In the trends of the $s$-process elements, Sgr bears similarities to Fornax and the LMC, but it differs from GES and, to an even greater extent, from Sculptor. This difference may be attributable to metallicity effects. Indeed, at lower [Fe/H], heavy element production is dominated by the $r$-process, while AGB stars (usually with $\rm{M}<4\rm{M_{\odot}}$) contribute to $s$-process enrichment with a time delay of $\sim 1$ Gyr after the onset of the star formation activity \citep{2004Travaglio}. This comparison also suggests that systems, such as Sgr, Fornax, and the LMC, that experienced prolonged periods of star formation allow AGB stars to significantly contribute to their chemical enrichment \citep{2010Letarte, 2022Carrillo}, in contrast to GES and Sculptor, which show less extended star formation. For the former systems, it is plausible that their masses were sufficient to retain gas, which was crucial for reaching the stage where AGB stars could play a substantial role in the enrichment process. In contrast, GES appears to have had a shorter star formation process \citep[extended up to $\sim 3$ Gyr,][]{2020Das,2024Ernandes} that ceased before AGB stars could influence its chemical composition, likely due to its early accretion into the MW, as also suggested by recent studies (e.g., \cite{2021Hasselquist, 2024Ernandes}).

Differences are also evident in the $r$-process abundance, where the Eu level in Sgr seems comparable to that of the MW but is less enhanced than in GES ($\sim 0.3$ dex difference) and Fornax and LMC ($\sim 0.2$ and $\sim 0.1$ dex respectively). Despite the limited data points for Sgr, we observe a clear downward trend in metallicity, similar to what is seen in the LMC and Fornax in comparable metallicity regimes.

In light of these considerations, we conclude that the gradual enrichment observed in the dSph Sgr, Fornax, Sculptor or the massive LMC contrasts with the more exotic and episodic events that could account for an earlier and higher levels of $r$-process pollution seen in GES. Sculptor exhibits only a mild enhancement in Ba, which may suggest that during its estimated 7 Gyr of SFH \citep{2012deboersc}, metal-poor AGB stars had sufficient time to enrich the system, albeit to a lower level compared to Sgr or Fornax. 

\subsection{Chemical enrichment histories revealed with heavy-elements} \label{sect:heavy}
\begin{figure*}[!h]
\centering
 \includegraphics[width=\textwidth]{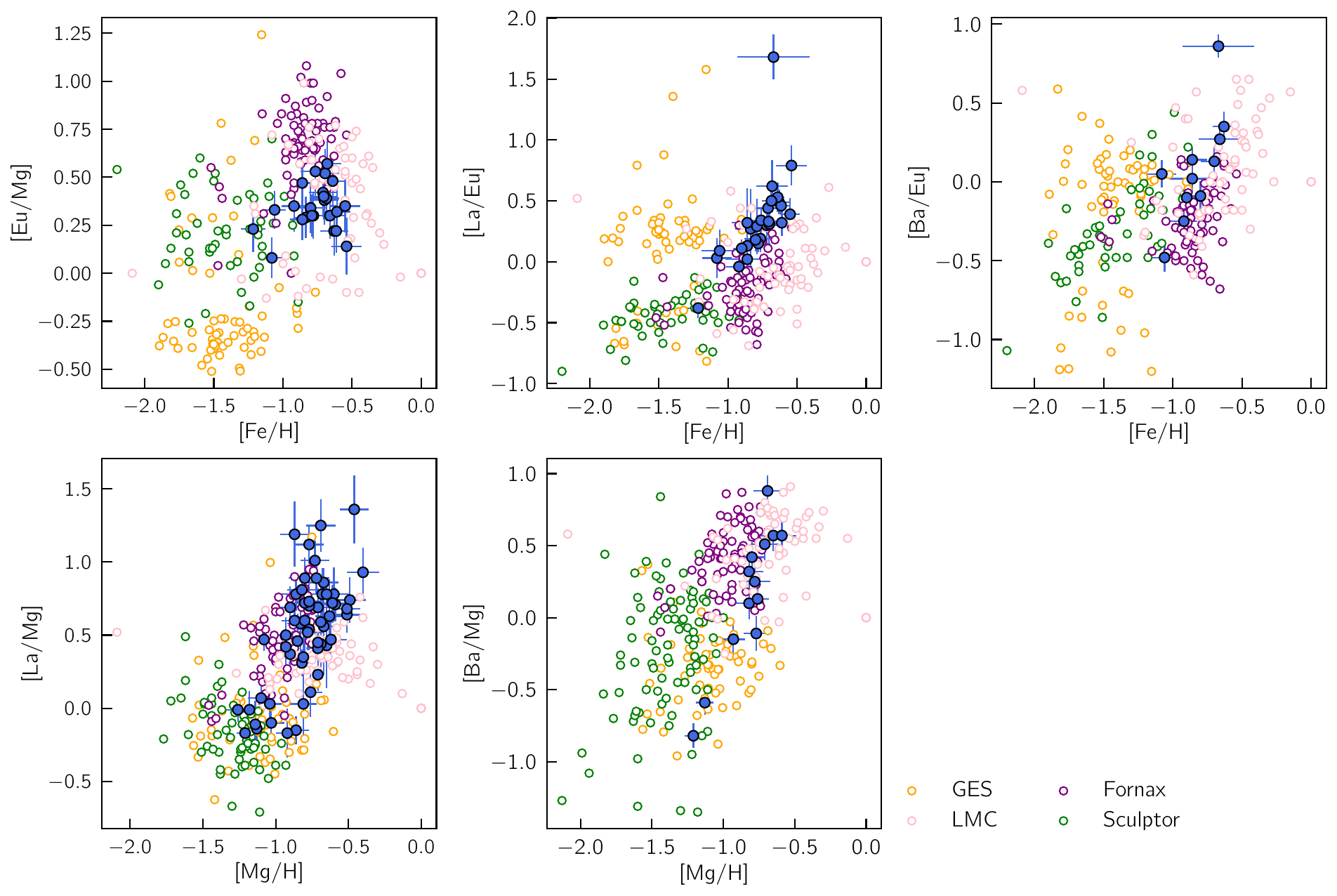} 
\caption{\small{Different chemical combinations of $n$-capture elements (La, Ba, Eu) and $\alpha$ element (Mg) for Sgr, compared to the systems shown in Fig~\ref{fig:chemical_trend_mw} (illustrated with the same colors). In the top row, the chemical ratios are examined as a function of metallicity, while in the bottom row as a function of Mg abundance.}}
 \label{fig:chemical_comb}
\end{figure*}

In Fig.~\ref{fig:chemical_comb} we explore five different chemical planes involving [$n/\alpha$]-elements and [$s/r$]-elements for Sgr, comparing them with the other satellites discussed in Sect. \ref{sect:satellites}. These chemical ratios are widely used in the Galactic Archaeology field as their decoupled origins are known to help trace the chemical evolution of the galaxy and provide clues to the various nucleosyntheis channels which have chemically enriched the system (e.g., \citealt{2008Sneden,2013Nomoto, 2019Skuladottir, 2020Reichert}).

The [Eu/Mg] ratio as a function of metallicity is a good proxy for estimating the contribution of delayed sources to Eu production, alongside prompt nucleosynthetic channels like SNe II, which produce both Eu and Mg \citep{2018Cote,2020Skuldelayed}.  As previously mentioned, these delayed sources have been linked to exotic scenarios involving NSMs or MR-SNe \citep{2018Cote, 2020Skuldelayed}. 
\cite{2024Sestito} interpreted the spread observed in [Eu/Mg] within a more limited metallicity range as evidence of a partial contribution from delayed $r$-process sources, such as compact binary mergers. However, it is important to consider that these scenarios may not necessarily be responsible for the increasing [Eu/Mg] trend. It is possible that in dSphs with prolonged SFHs, another source contributing to the rise in $r$-process levels could involve massive stars formed during a later period of star formation, while gas was still available in the galaxy \citep{2021Reichert}.

In Sgr the [Eu/Mg] trend and supersolar enrichment has been previously suggested in the literature \citep{2013McWilliam,2019Skuladottir,2020Skuldelayed,2020Reichert} and analyzed by \cite{2024Sestito} in the VMP regime of Sgr. Our PIGS-HR-Sgr data confirm an increase in [Eu/Mg], which appears to be consistent with more massive systems that experienced prolonged star formation SFHs, though much less pronounced than in Fornax and LMC. Similarly, Sculptor, despite occupying a different metallicity regime, exhibits a comparable rise in [Eu/Mg] to that observed in Sgr. While both systems likely experienced star SFHs long enough to allow delayed $r$-process sources—estimated by \cite{2020Skuldelayed} to require $\gtrsim 4$ Gyr to enrich the ISM, attributing this enrichment solely to such exotic scenarios is not straightforward.

For GES, although the data from \cite{2022Carrillo} do not reveal a clear trend, the work of \cite{2021Matsuno} identified an $r$-process enrichment linked to NSMs with a delay time contribution. However, compared to systems like Sgr, Fornax, and the LMC, \cite{2021Matsuno} highlighted a distinct difference in the extent of this $r$-enrichment. In fact, the weaker increasing trends in this chemical space for GES may suggest a shorter early star formation period, which is thought to be truncated $\sim$ 9-10 Gyr following $\sim 3$ Gyr of star formation \citep{2020Belokurov, 2020Das,2023Giribaldi,2024Ernandes}. This would imply that a source with a delay time of about 4 Gyr, as proposed by \cite{2020Skuldelayed}, may not have had sufficient time to enrich GES as significantly as other massive systems (Sgr, Fornax, LMC). Alternatively, it is more likely that the violently truncated star formation in GES prevented a more prolonged SFH with subsequent star formation episodes which would rise the $r$-process level in the galaxy.

\cite{2013McWilliam}, in explaining the [Eu/Mg] trend in Sgr, suggested that a top-light IMF, lacking the more massive SNe II that contribute hydrostatic elements like Mg to the ISM, played a role. From this perspective, the difference observed in GES could also suggest that CCSNe continued to enrich this system in Mg to higher [Fe/H] values, unlike what happened in Sgr. This implies that the IMF of GES may have been richer in more massive stars, but that star formation was quenched earlier due to its violent infall into the MW. \cite{2004Shetrone} had already interpreted the rising [Eu/Ca] trends found not as being due to $s$-process sources, but rather as arising from sources different from those responsible for the $\alpha$ abundances.

Without need of advocating to the IMF, the differences between GES and Sgr can be explained by the fact that this latter might have had more time to continue forming stars before its infall into the MW, allowing for a later star-forming episodes to increase its Eu levels. It is also possible that gas stripping was less effective in the central regions of the main body of Sgr, preventing a complete quenching of star formation. This could have left residual gas that was exhausted only later, after the galaxy's first pericentric passage \citep[$\sim 5$ Gyr][]{2020Ruiz}. Moreover, due to the spread in the Eu measurements in our analysis, it is not possible to rule out the contribution to the $r$-process enrichment from later massive stars and their subsequent SNe explosions. Disentangling these scenarios in Sgr is challenging. However, considering the overall analysis, we believe that rise in the [Eu/Mg] panel most likely reflects the extended SFH of Sgr, characterized by multiple star-forming episodes that enriched the ISM in both $r$- and $\alpha$ elements without producing a pronounced [Eu/Mg] trend. We also examined the Eu abundances as a function of age, but the resulting trend did not provide a clear timescale to constrain the timing of Eu enrichment. 

The second and third panels of the top row in Fig.~\ref{fig:chemical_comb} provide valuable diagnostics for examining the contribution of the $s$-process channel. As noted in the study by \cite{2024Sestito}, Sgr exhibits an increasing trend of approximately 1 dex in [Ba/Eu] versus [Fe/H], a pattern also observed in the [La/Eu] ratio. This upward trend suggests a substantial contribution from s-process production channels relative to the $r$-process  ($\sim 96 \%$ of Eu is produced through $r$-process \citeauthor{2004Travaglio}, \citeyear{2004Travaglio}, \citeauthor{2014Bisterzo}, \citeyear{2014Bisterzo}). This reinforces evidence for a significant AGB star contribution in Sgr and indicates a combination of ($r+s$) formation sites, as also proposed by \cite{2018Hansen} for the metal-poor sample analyzed in their study. This tendency is also observed in the LMC and Fornax, though it is less pronounced and more gradual in GES. This suggests that the progenitor of GES was dominated by different nucleosynthetic processes compared to these other systems. By combining the information from these panels with that of the first, we can conclude that the dSphs Sgr, Fornax, and the LMC show enrichment in heavy elements from both prompt and delayed sources of $r$-process elements, balanced by a delayed onset of AGB star contributions \citep[as found by][]{2010Letarte, 2014Lemasle, 2018Hansen, 2020Reichert}, beginning around [Fe/H] $\sim -1.2$ for Sgr, similar to its $\alpha$-knee, indicating a similar time delay (the same behavior is shown by Sculptor where this phenomenon happens around -2.0 \citeauthor{2019Hill}\citeyear{2019Hill}). In contrast, GES experienced intense early enrichment in Eu with a smaller $s$-process contribution \citep[as also evinced by][]{2021Matsuno}, likely due to a shorter and more abruptly quenched SFH. Finally, Sculptor exhibits a very high initial Eu abundance, which might be caused by compact merger events, followed by a steep rise in Ba at higher metallicities which can witness the $s$-enrichment from metal-poor AGBs.

The trends illustrated in the last two panels of the bottom rows serve as further tools for assessing the contribution of AGB stars in the closed system of Sgr. The rise in [La/Mg] of $\sim 2$ dex excludes a scenario where only the yields from SNe II contribute to the enrichment of the system, as this would result in a flat [La/Mg] or [Ba/Mg] distribution \citep[e.g.,][]{2008Pignatari,2021Cowan}. This is because the $s$-processes occurring in AGB stars increase the levels of La and Ba relative to Mg, which is primarily produced by SNe II \citep{2008Pignatari, 2015Cristallo}. Typically, this pattern is associated with the mass of these systems, which is high enough to retain their gas, allowing them to evolve to the point where AGB stars become the dominant contributors. As shown in Fig~\ref{fig:chemical_comb} the observations in Fornax and the LMC exhibit similar behaviors, which are supported by the literature \citep{2008Pompeia,2010Letarte,2020Reichert}. Sculptor also demonstrates a rapid increase in the [Ba/Mg] vs. [Mg/H] relation; however, the behavior for La is less distinct due to its more even distribution between $s$- and $r$-processes (with 75\% of La produced by the $s$-process \citeauthor{2014Bisterzo}, \citeyear{2014Bisterzo}) contrary to Ba (85 \% produced by $s$-process \citeauthor{2014Bisterzo}, \citeyear{2014Bisterzo}). \cite{2019Hill} also discussed the potential biases in La measurements, as they are derived from weak lines. The GES data show a similar trend, where the increase in Ba may indicate an additional contribution from $s$-process elements occurring on timescales comparable to the onset SN~Ia. \cite{2024Ernandes} found similar results and concluded that the absence of an increasing [Ba/Mg] ratio at higher metallicities (in our case, [Mg/H]) is indicative of a significant halt in star formation in GES, which can also lead to a lack of substantial $s$-process contribution as metallicity increases. Due to the limited Ba measurements in Sgr, it is challenging to discern this difference in their trends. The low ratios of [La/Mg] and [Ba/Mg] at lower values of [Mg/H] in Sgr reflect a period of strong initial star formation in the galaxy, characterized by a rapid contribution from CCSNe (Mg) compared to the delayed yields from AGB stars (Ba). Hence, even if CCSNe occur at later times from the remaining gas in Sgr (before its complete quenching), contributing to delayed $r$-process enrichment and $\alpha$ elements, the stronger contribution of $s$-process elements from AGB stars counteracts this effect.

Without measurements of the ratio of heavy ($hs$, e.g., La, Ba) to light ($ls$, e.g., Sr, Y, Zr) $n$-capture elements, constraining the mass of the AGB stars that enriched the ISM of Sgr becomes challenging. The [($ls$)/($hs$)] ratio is crucial for this purpose \citep{2004Travaglio, 2010Pignatari}. Based on the findings of \cite{2005McWilliam} and \cite{2013McWilliam}, MP and low-mass AGB ($\lesssim 2 \rm{M_{\odot}}$) are invoked to explain the chemical properties of Sgr. \cite{2004Shetrone} also appeals to an enhanced $r$-process contribution to explain the higher [La/Y] ratios measured for Sgr in comparison with the MW, as La originates from both $r$-process and $s$-process channels. A different view is given by \cite{2018Hansen}, who find that their best fit for the chemical abundances of their VMP Sgr sample is provided by intermediate-mass AGBs ($\sim 5 \rm{M_{\odot}}$). 

\subsection{Age-metallicity relation}\label{sect:AMR}

The SFH of Sgr has been deeply investigated over the years \citep[e.g.,][]{1999BellazziniI,1999Bellazzini,2020Ruiz}. The star formation episodes have been dated and compared them with the kinematic and orbital history of this dwarf galaxy, which has been modeled in various works \citep[e.g.,][]{2010Penarrubia,Vasiliev20,2024Davies}. In this context, accurate age estimates are crucial for constructing a detailed evolutionary history of the system. However, deriving this information is not without its challenges due to the indirect methods and techniques used for age determination. Using isochrone techniques in combination with deep-band photometry, \cite{2000Layden} provided a detailed SFH for both M54 and Sgr field stars. Their study identified multiple episodes of star formation over a span of 0.5 to 14 Gyr, concluding a protracted SFH, a finding later validated by \cite{Siegel07}. Through photometric data from the HST and employing isochrone and CMD fitting, \cite{Siegel07} identified several stellar populations within the main core, ranging from an old MP population ($\sim$10-12 Gyr) to a very young population of only a few Myr. Stellar clusters are often used to determine ages more easily. 

For the PIGS-HR-Sgr sample we derived ages following Povick et al. in prep. Using a finely spaced grid of PARSEC \citep{girardi02padova,bressan2012parsec,marigo2017} isochrones, all points within 7$\sigma$ of each star are selected in a 10-dimensional parameter space, which includes optical/NIR multiband photometry from Gaia DR3 \citep{gaia2016mission,gaia2022dr3} and 2MASS \citep{skrutskie20062mass} as well as Teff, [M/H], and $\log{g}$. For each isochrone point, the $\chi^2$ value is calculated incorporating an additional term that functions as a prior based on the Kroupa IMF \citep{kroupa2001imf,kroupa2002imf}. The age, extinction, and mass values for each stars are then obtained with a weighted mean of the isochrone points, with the weights determined by the modified $\chi^2$ values. 
\begin{figure}[t]
\centering
 \includegraphics[width=0.97\columnwidth]{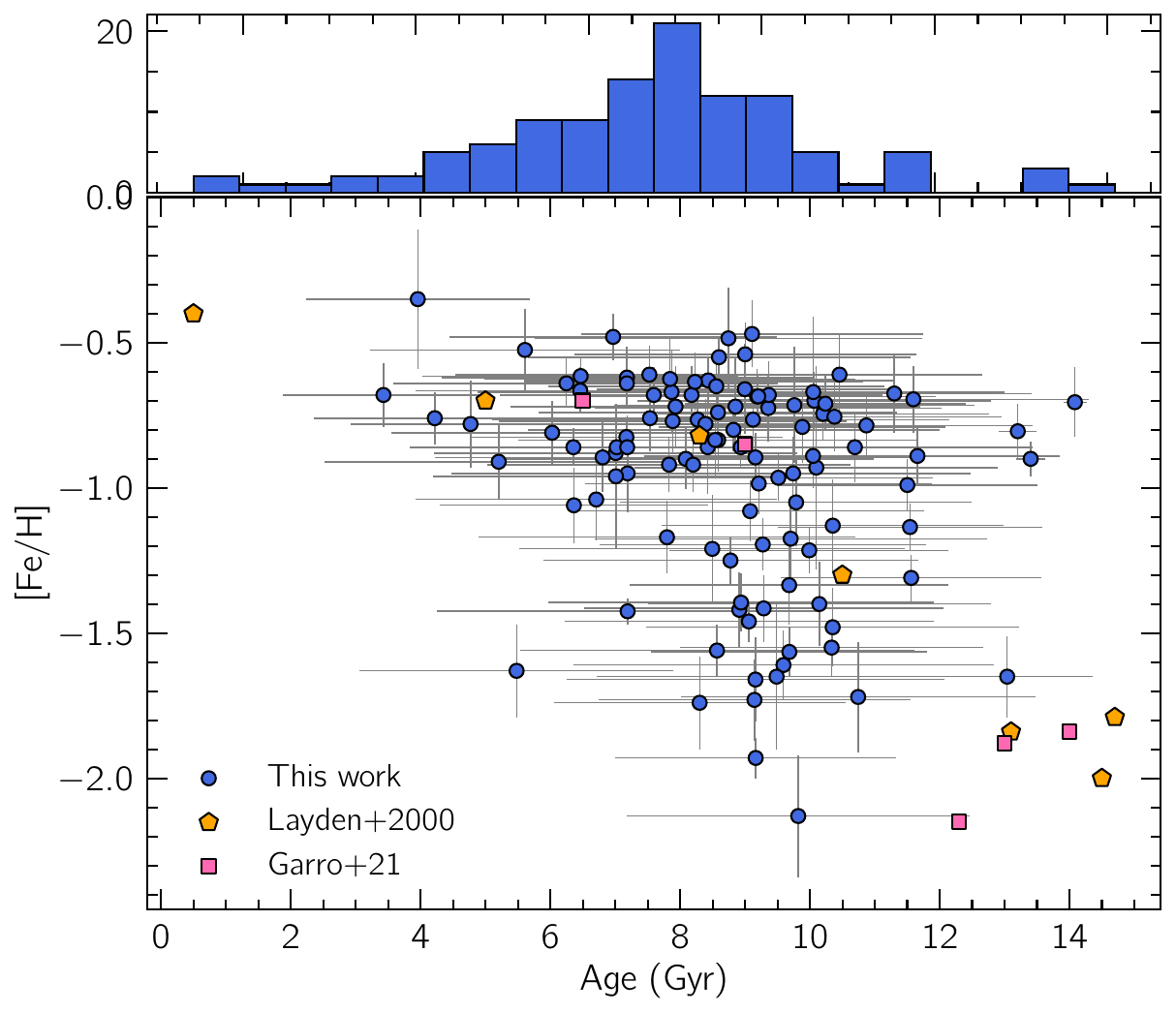} 
\caption{\small{Age-metallicity relation for the Sgr stellar sample (blue circles). The pink dots indicate five GCs associated to the Sgr streams reported by \cite{2021Garro}, while the orange pentagons represent both GCs and the Sgr field population derived with isochrone fitting by \cite{2000Layden}.}}
 \label{fig:AMR}
\end{figure}

When deriving ages for the stars, their {[$\alpha$/Fe]} values were fixed to solar. It was found that incorporating the measured {[Mg/Fe]} as a proxy for overall {[$\alpha$/Fe]} resulted in many more spurious ages, particularly for $\alpha$-enhanced stars, leading to ages less consistent with a dominant population of $\sim$10~Gyr as found by \cite{layden1997sgr}. The uncertainties in the abundance measurements also contribute to the scatter observed when using $\alpha$ abundances. Furthermore, the vast majority of stars in this sample have [Mg/Fe]$\sim$0.0 dex, so assuming solar [$\alpha$/Fe] was determined to be reasonable. 

The resulting AMR for our targets (blue circles) appears in Fig.~\ref{fig:AMR}. To guide the eyes, we also present the age estimates from \cite{2000Layden} for the four GCs associated with the Sgr remnant (NGC 6715/M54, Terzan 7, Terzan 8, and Arp 2) as well as for three field populations. Additionally, we report the ages inferred by \cite{2021Garro} for five more GCs recently identified as part of the Sgr streams \citep{2020Bellazzini}. Despite the large scatter in our AMR, a general agreement in the trends can still be observed across different studies. The PIGS-HR-Sgr AMR is fairly flat for ages between 6 and 10 Gyr, with a more pronounced increase in metallicity evident among younger stars. The peak at $\sim$ 6-8 Gyr in our data corresponds to an intermediate-age population with average [Fe/H] $\sim -0.85$. This population has been previously identified as having undergone intense and rapid star formation activity and constitutes the predominant population \citep{1999Bellazzini,Siegel07}. 
Stars in this population likely formed before Sgr's first infall into the MW's potential ($\sim 5$ Gyr ago; \citealt{2020Ruiz}), an event that may have triggered a burst of star formation in the MW \citep{2020Ruiz} and likely led to the subsequent quenching of star formation in Sgr due to gas stripping, along with a natural decline in the dwarf's SFH.

However, as noted in other studies, residual star formation activity can be identified with stars younger than 6 Gyr. These residual star-forming episodes could be related to interactions with the MW. While ram-pressure stripping likely removed gas from Sgr, it may also have compressed the remaining gas, triggering new star formation episodes—a scenario explored in the recent work by \cite{2024Yang} in other dSphs, but not in Sgr. Supporting evidence for these later star formation activities comes from our identification of a few younger stars with ages of $\sim$4 Gyr, likely reflecting the lower metallicity limit of our PIGS-HR-Sgr sample ([Fe/H] $\gtrsim -0.5$ dex). The presence of younger populations with ages of $\sim$3–4 Gyr further supports the idea that, despite gas quenching, Sgr retained enough gas in its central regions to form stars from its remaining gas reservoir.

Looking at the older stars ($> 8$ Gyr) there is considerable scatter in metallicity between $\sim 8-10$ Gyr, which corresponds to the age range with the largest uncertainties ($\sim 2.5-3$ Gyr). We can assume that this old stellar populations may have potentially originated from the gas of the nuclear center of the galaxy and later displaced during Sgr's orbital history. Three stars with age $> 12$ Gyr show surprisingly high metallicity values (i.e., $-0.5 \lesssim \rm{[Fe/H]} \lesssim -1.0$) and low age uncertainties ($\lesssim 0.3$ Gyr). We reviewed the input atmospheric parameters and metallicities and found no anomalies. It appears that the three stars are approaching the isochrone grid edge in age. Additionally, the density of isochrone points in metallicity,$ \teff$, and $\logg$ space drops near the three outliers. These two factors most likely explain what is happening with these three stars. 

Examining the other outputs from the isochrone fitting, we verified that the extinction values agree with those used to derive the $\rm{T_{eff}}$ values and that the extinction remains relatively uniform across the spatial region covered by the Sgr targets. The corresponding plots are presented in Appendix \ref{sec:append_ages}, along with the range of resulting masses, which fall within the typical mass range of low-mass RGB stars ($\sim 0.8-2.0~\rm{M_{\odot}}$), consistent with older and intermediate-age stellar populations.

Studies such as the ones from \cite{2009Revaz,2012Revaz} demonstrate that the star formation in massive dSphs (above $\sim 3 \times 10^{8},\rm{M_{\odot}}$) is naturally linked to the mass of the system. In such cases, star formation occurs in more homogeneous and less complex scenarios, where external factors such as gas removal or ram pressure are not necessarily required or would leave a noticeable impact if present. This contrasts with lower-mass dwarf galaxies, which exhibit more diverse SFHs that are strongly influenced by quenching mechanisms, including SNe feedback or external processes. For Sgr, its mass suggests a reduced impact from such mechanisms. Indeed, despite its merging history with the MW, the AMR observed in our central PIGS-HR-Sgr sample indicates that star formation has not been entirely halted. Consequently, the overall shape of its AMR likely reflects its intrinsic SFH, which is closely linked to its initial mass.
 
Lastly, due to the scatter in ages around $\sim 8$–10 Gyr, it is not trivial to determine the age at which our estimate of the knee occurred. In the study by \cite{2014deBoer} on the Sgr stellar stream, the knee was constrained to an age of $\sim 11$ Gyr. Based on our value of $\rm{[Fe/H]}_{\rm{knee}} = -1.05$, the knee appears to occur at around $\sim 8$–9 Gyr, which is slightly later than the age found by \cite{2014deBoer}. This discrepancy is reasonable, as they reported a lower metallicity for the knee, [Fe/H] = -1.27~dex, which was calculated for stream stars using different methodologies. Nevertheless, considering the timescales in both studies, this would place the occurrence of the knee around 1-3 Gyr after the formation of the system, as also found in \cite{2015deboer}. This timescale is consistent with the delay in the onset of Type Ia supernovae \citep{2004Strolger,2006Sullivan,2021Wiseman}. 

\section{Summary and conclusions} \label{sect:conclusion}

The history and chemical evolution of the interacting Sagittarius dwarf galaxy has been studied using both photometric and spectroscopic techniques. In this work, we assemble a large high-resolution dataset covering $-2.13 < \rm{[Fe/H]} < -0.5$~dex, which represents the largest optical high-resolution dataset with measurements of 13 chemical species over this metallicity range. Thanks to a prior selection based on metallicity using PIGS photometry, we conducted a spectroscopic follow-up of 111 Sagittarius stars with FLAMES/GIRAFFE. By inferring 14 chemical elements with an average uncertainty of $\sim 0.09$ dex, we investigated the chemical enrichment history of this galaxy, placing it in the context of its interaction with the MW by comparing it to other satellites and structures. 

By inferring RVs for the entire stellar sample  (see Fig.~\ref{fig:RVhisto}) and finding a good match with the distribution from the literature, we can confirm the Sgr membership selection performed using \textit{Gaia} astrometry. Although we observe a slightly higher proportion of MR stars, particularly with $\rm{[Fe/H]} > -1.0$ (see Fig.~\ref{fig:CCD}), the resulting spectroscopic metallicities confirm the effectiveness of the PIGS photometry in preselecting candidates based on their iron content. This leads to a [Fe/H] coverage ranging from -2.13 to -0.35, and peaking at [Fe/H] $\sim -0.68$.

We compared the various [X/Fe] trends explored in our analysis with different intermediate and high-resolution samples of the Sgr core from the literature (Fig.~\ref{fig:chemical_trend}). Our Sgr sample aligns with the expected Galactic chemical trends: $\alpha$ elements decrease with increasing metallicity, while $n$-capture elements exhibit the opposite trend, which reflects their distinct astrophysical production sites. This pattern is also evident for the two Fe-peak elements, whose enhancements, though at different levels, are influenced mostly by SN~Ia in the case of Co I and by both SN~Ia and CC-SNe for CrI, which leads to different trends with [Fe/H].

This chemical investigation was extended to compare the chemical trends of the MW with those of its most massive satellite systems (Fig.~\ref{fig:chemical_trend_mw}). Consistent with the literature \citep{2022Carrillo}, we found that the satellites are deficient in hydrostatic elements relative to the MW, a difference potentially linked to a less efficient early star formation period, or connected as in \cite{2013McWilliam} and \cite{2017Hasselquist} to a top-light IMF that lacks the most massive stars. However, we acknowledge that the observed differences in elements such as Mg and Ca are challenging to disentangle, as they may arise from the system's star formation rate or the yields of SN~Ia. Further investigation, including the use of chemical evolution models, is therefore needed to confirm the true nature of Sgr's IMF.

Trends in Fe-peak and odd-Z elements appear similar among the more massive systems (Sgr, LMC, GES, and Fornax) and the MW. However, an insightful distinction arises with the $n$-capture elements, especially the $s$-process La and Ba, which exhibit a marked enhancement in Sgr compared to the MW—a trend that aligns with Fornax and the LMC but diverges from the GES. The latter also shows an enhanced $r$-process abundance compared to the MW and the other systems, among which Fornax and the LMC follow GES. Taking into account the [Fe/H] dependence, which affects the trends of the systems that cover different metallicity ranges, it is yet possible to conclude that the dSph Sgr, Fornax, and the LMC show a larger contribution from AGB stars (possibly compared to SNe) relative to the MW. This result has been found in many studies concerning dSph galaxies \citep[e.g.,][]{2013McWilliam,2010Letarte,2021Hasselquist}. This is less true for GES, which is expected since this system only evolved over the course of 2–3 Gyr, a time frame insufficient for AGB stars to enrich the system. However, AGB contributions have been found to be part of its evolution \citep{2019Hill, 2024Ernandes} hence, it is possible that the lower values of Ba and La reflect a metallicity effect on the AGB yields.

A more in-depth investigation of the chemical enrichment histories of these MW satellites was carried out throughout the exploration of heavy elements, as shown in Fig.~\ref{fig:chemical_comb}. Sgr and Sculptor show a mild increase in the [Eu/Mg] trend, much less pronounced than in Fornax and the LMC. A stronger increase in [Eu/Mg] may suggest extended SFHs with contributions from delayed $r$-process sources, such as NSM \citep{2020Skuldelayed}. These appear to be absent or not prominent in Sculptor and Sgr, and occur on timescales of $\lesssim 4$ Gyr in GES, according to \cite{2021Matsuno}. The most evident difference is between systems with long SFHs, such as Sgr (or Fornax and the LMC), compared to GES, which had halted and truncated star formation but, by contrast, an IMF rich in massive stars that continued to enrich in Mg to higher metallicities.

Looking at the $[s/r]$ ratio versus [Fe/H] (i.e., [La/Eu] and [Ba/Eu]), Sgr shows an increasing trend of $\approx 1$ dex, slightly higher than that of Fornax and the LMC, which suggests enrichment in heavy elements due to a delayed onset of AGB stars, balancing both prompt and delayed sources of $r$-process elements \citep[as noted in previous studies, e.g., ][]{2010Letarte,2018Hansen}. Both GES and Sculptor exhibit a high level of $r$-process enrichment; however, it remains steady in GES, which indicates a halted SFH, while it increases in Sculptor, which reflects ongoing AGB contributions. This information was further supported by the study of $[s/\alpha]$ ratios, which confirms a significant contribution from AGB stars in Sgr beginning at higher [Mg/H]. This suggests that early Mg production was driven by CCSNe, and later superseded by delayed yields from AGB stars, a pattern that is also observed in the LMC and Fornax. In conclusion, even without detailed chemical evolution models, these chemical trends related to $n$ capture serve as powerful diagnostics for understanding the evolution of these systems, being closely tied to their SFH, as evinced in our analysis.

We derived $\rm{[Fe/H]_{\alpha-knee}} = -1.05$ for Sgr (see Fig.~\ref{fig:knee}), and incorporated literature datasets \citep[e.g.,][]{2005Monaco,2007Sbordone,2017Mucciarelli}. Despite a nonhomogeneous metallicity coverage, the occurrence of the knee suggests that Sgr experienced slower early enrichment compared to the MW, but was still able to utilize its gas to continue star formation to higher [Fe/H] values than the LMC, which followed an isolated evolution. To answer a question already raised by \cite{2024Sestito}, no strong evidence of SN~Ia was found at [Fe/H] $\lesssim$ -1.0. Relying on the measured $\alpha$ abundances, we examined the ratios between hydrostatic (Mg) and explosive elements (Ca and Ti) in Fig.~\ref{fig:delta_alpha}, and propose the presence of HNe enriching the more metal-poor stars in Sgr, as also suggested by \cite{2024Sestito}.

Finally, the AMR presented in Fig.~\ref{fig:AMR} aligns with the photometric studies of the SFH by \cite{2000Layden} and \cite{Siegel07}. Despite significant scatter, we identify the main populations within Sgr: an older, metal-poor population aged 8-10 Gyr, which is also the period during which the knee can be placed. This is followed by a dominant intermediate-age population peaking around 6-8 Gyr, likely formed before Sgr’s first infall onto the MW during a strong and rapid star formation period. Following this, Sgr continued to form stars by compressing the remaining gas, as evidenced by stars with ages $\sim 3-5$ Gyr, despite the effects of gas stripping. However, our sample does not include the most metal-rich populations ([Fe/H] $> -0.5$) or the youngest stars ($< 3$ Gyr).

To summarize,  we propose that Sgr exhibits a chemical evolution characteristic of a massive dwarf galaxy. Its stellar populations, sampled from the galaxy's central regions and thus largely unaffected by contamination from MW gas, reflect Sgr's independent chemical enrichment history. This history includes multiple episodes of star formation, peaking within the first few billion years after its formation and subsequently declining. The results for heavier elements reveal an enrichment in $n$-capture elements, which is indicative of a prolonged SFH, further supported by the AMR. The various chemical trends suggest contributions from different enrichment sources, including various types of SNe and AGBs, while showing no evidence of chemical sequences attributable to MW gas infall. This is expected, as few of our targets are likely to have formed after the first interaction with the MW. Moreover, the target field is located in a region of the galaxy where MW gas infall is highly unlikely. When comparing Sgr with other significant satellites of the MW and its halo, we observe that many of its chemical trends resemble those of dwarf galaxies capable of sustaining a longer SFH.

Despite the diagnostic power of chemical abundances, in future work, we aim to use chemical evolution models to further disentangle the history of the core of Sgr and constrain the contributions of SNe across different timescales. With precise and accurate models, the chemical evolution and SFH can be better connected and used to trace the precise history of this galaxy, from its progenitor through the long interaction period with our Galaxy.

Additionally, upcoming spectroscopic surveys like 4DWARFS \citep{20234dwarf} or MOONS \citep{2020moons} will supply more data to enhance our study of the Sgr stellar populations within both the core and the stream. These surveys will allow us to investigate their common progenitor, trace the SFH, and conduct chemodynamical studies, all of which will expand our understanding of Sgr as a satellite of the MW. 

\section*{Data Availability}
Tables \ref{tab:lines} and \ref{tab:AP} are only available in electronic form at the CDS via anonymous ftp to \url{https://cdsarc.u-strasbg.fr/} \url{(130.79.128.5)} or via \url{http://cdsweb.u-strasbg.fr/cgi-bin/qcat?J/A+A/}.

\begin{acknowledgements}
SV thanks ANID (Beca Doctorado Nacional, folio 21220489) and Universidad Diego Portales for the financial support provided. SV and PJ acknowledge the Millennium Nucleus ERIS (ERIS NCN2021017) and FONDECYT (Regular number 1231057) for the funding. A. R. A. acknowledges support from DICYT through grant 062319RA. ES acknowledges funding through VIDI grant "Pushing Galactic Archaeology to its limits" (with project number VI.Vidi.193.093) which is funded by the Dutch Research Council (NWO). This research has been partially funded from a Spinoza award by NWO (SPI 78-411).

This work has made use of data from the European Space Agency (ESA) mission {\it Gaia} (\url{https://www.cosmos.esa.int/gaia}), processed by the {\it Gaia} Data Processing and Analysis Consortium (DPAC, \url{https://www.cosmos.esa.int/web/gaia/dpac/consortium}). Funding for the DPAC has been provided by national institutions, in particular the institutions participating in the {\it Gaia} Multilateral Agreement.\\
SV led the analysis and various discussions in this work, contributed to writing most of the draft, and created all the figures presented in this work. AAR assisted with data reduction, analysis, and discussions related to the interpretation of the results. PJ and FS contributed significantly to the interpretation of the draft and the discussion of the results. JP helped revise the draft and provided the results for the stellar ages. AAA contributed the PIGS photometric data, which were used for target selection and largely contributed to the scientific discussion. VH and EF assisted with revising the draft, preparing the proposal, and setting up the OBs. PJ, NM, ES, and DA provided insightful scientific and editorial comments on the manuscript.\\
We thank the anonymous referee for their constructive comments and suggestions, which helped improve the quality and clarity of this manuscript.
\end{acknowledgements}

\bibliographystyle{aa}
\bibliography{bib}

\begin{appendix}\label{sec:append}
\onecolumn
\section{Radial velocities}\label{sec:append_rv}
\begin{figure*}[h]
\centering
 \includegraphics[width=0.75\columnwidth]{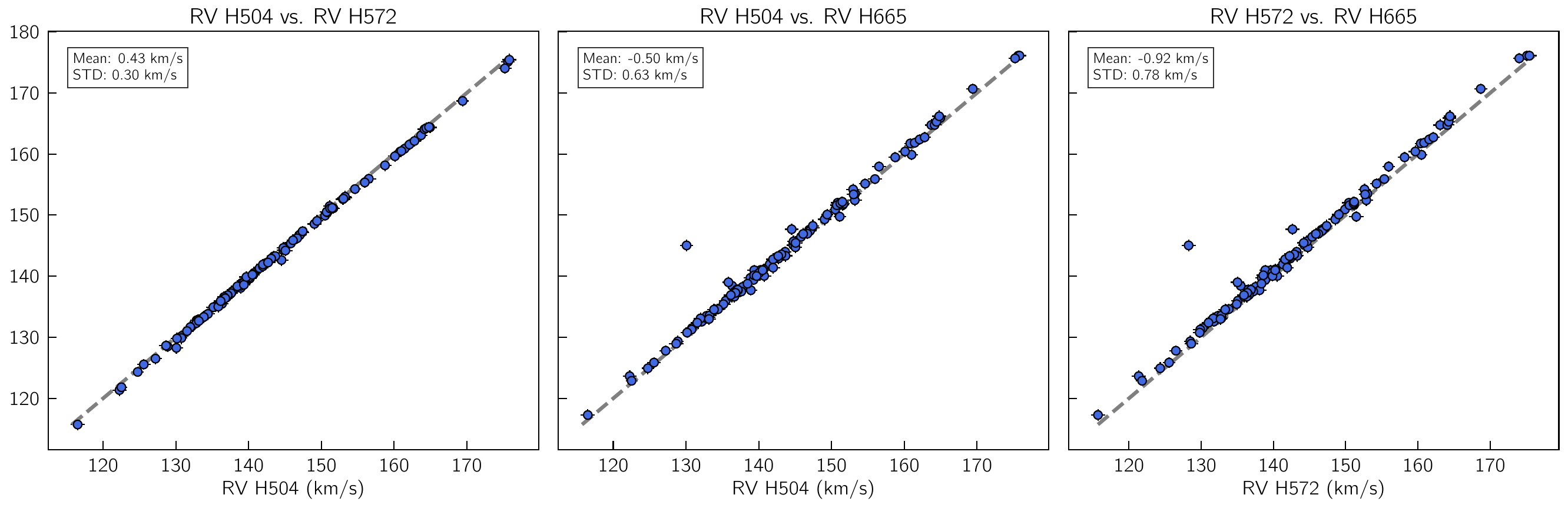} 
\caption{\small{Radial velocity values with their uncertainties for the entire FLAMES sample inferred using the three setups. A diagonal line is shown as a guide to better illustrate the agreement between the results of the different setups, which almost all fall on a 1:1 relation. For each panel, we report the mean offset and dispersion values for the different setups.}}
 \label{fig:rv_diff}
\end{figure*}

In Fig.~\ref{fig:rv_diff}, we present the RV values obtained using the method described in Sect.~\ref{sect:RV} for the three setups. Notably, the different sets of values show good agreement with the exception of two stars where the redder setup H665 yields RV values $\approx 5\,\mathrm{km\,s^{-1}} $ higher than the other setups. Upon inspecting the spectra of these two stars, we found that the spectra quality for the redder setup was compromised due to sky emission lines in this wavelength region. Consequently, we decided to compute the final RV for these two targets excluding the H665-RV values.
 
\section{Metallicities}\label{sec:append_met}
\begin{figure*}[h]
\centering
 \includegraphics[width=0.29\columnwidth]{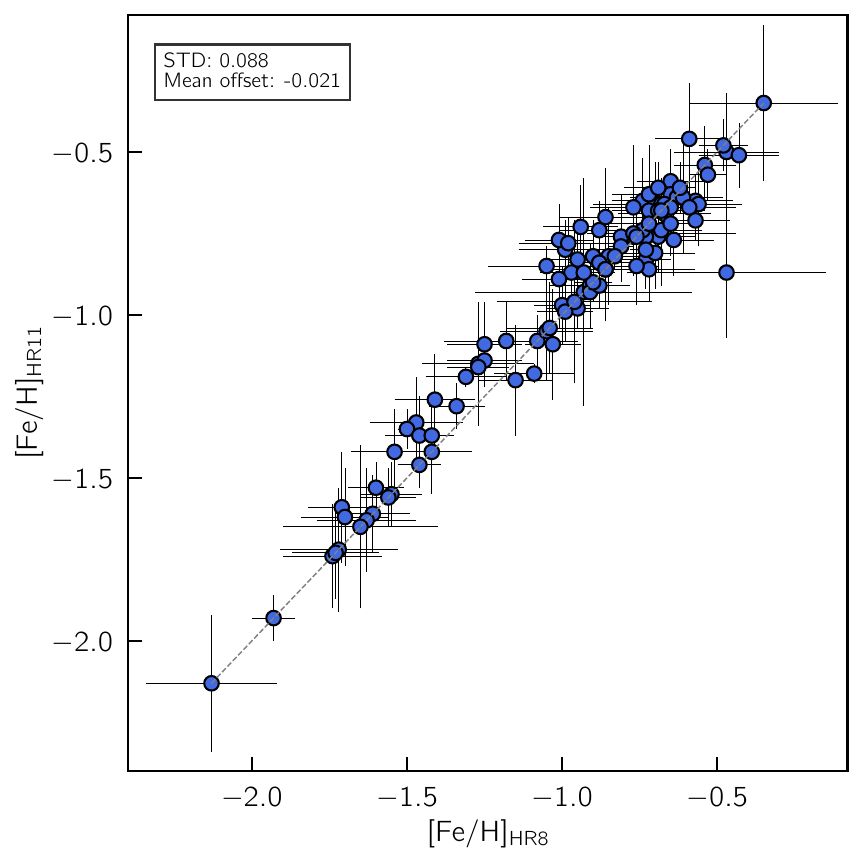} 
  \includegraphics[width=0.29\columnwidth]{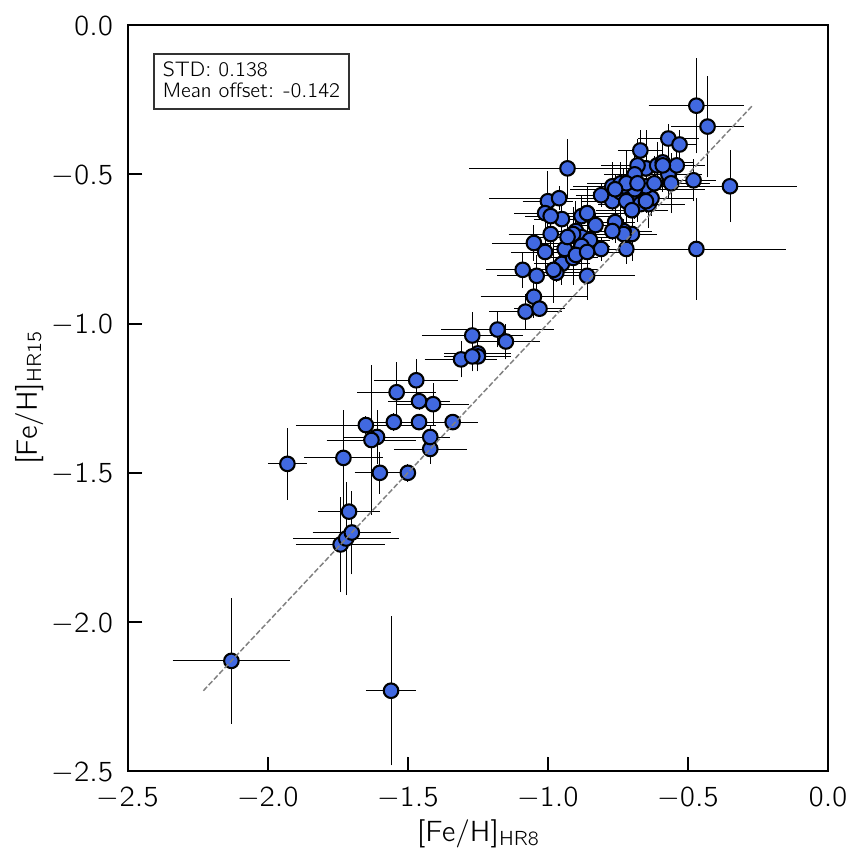}
  \includegraphics[width=0.29\columnwidth]{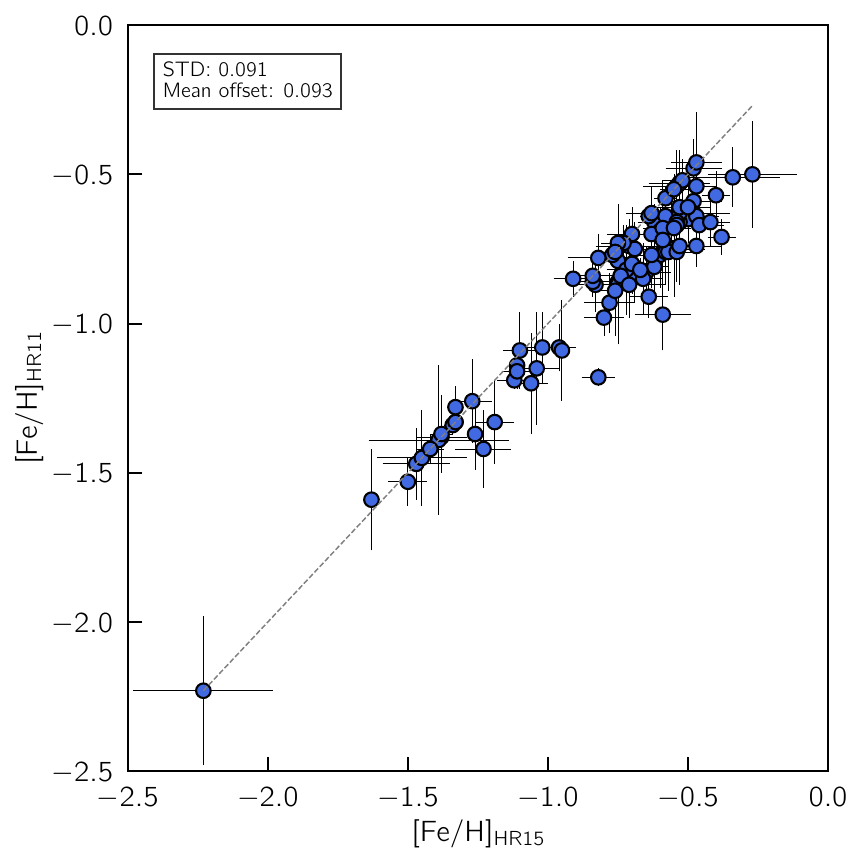} 
\caption{\small{Comparison of spectroscopic iron abundances obtained with the three different GIRAFFE filters, with the corresponding mean offset and dispersion values reported in each panel.}}
 \label{fig:met_comp}
\end{figure*}

The metallicity results from the three different setups are illustrated in Fig.~\ref{fig:met_comp}. Each metallicity sample was inferred following the methodology outlined in Sect.~\ref{sect:chemistry}. The left panel displays the [Fe/H] values selected for averaging as the final measurements for this study. The other two panels show that the metallicities obtained from HR15 are systematically higher compared to those from the other two setups. Therefore, the HR15 data were excluded from the final averaged iron abundances.

\section{$\alpha$-knee in NLTE}\label{sec:knee_nlte}

To determine the effects of NLTE abundance corrections on the derivation of the knee metallicity, we repeat the procedure described in Sect.~\ref{sec:knee}, considering only Mg abundances in both the original LTE results and the NLTE-corrected ones. The latter are obtained by applying corrections to the Mg line at 571.1 nm \citep{2017Bergemann}, as queried from the MPIA dataset\footnote{\url{https://nlte.mpia.de/}}.
We made use of a public python script\footnote{\url{https://github.com/anyadovgal/NLTE-correction}} which helps in automatizing NLTE corrections.
\begin{figure}[h]
\centering
 \includegraphics[width=0.38\columnwidth]{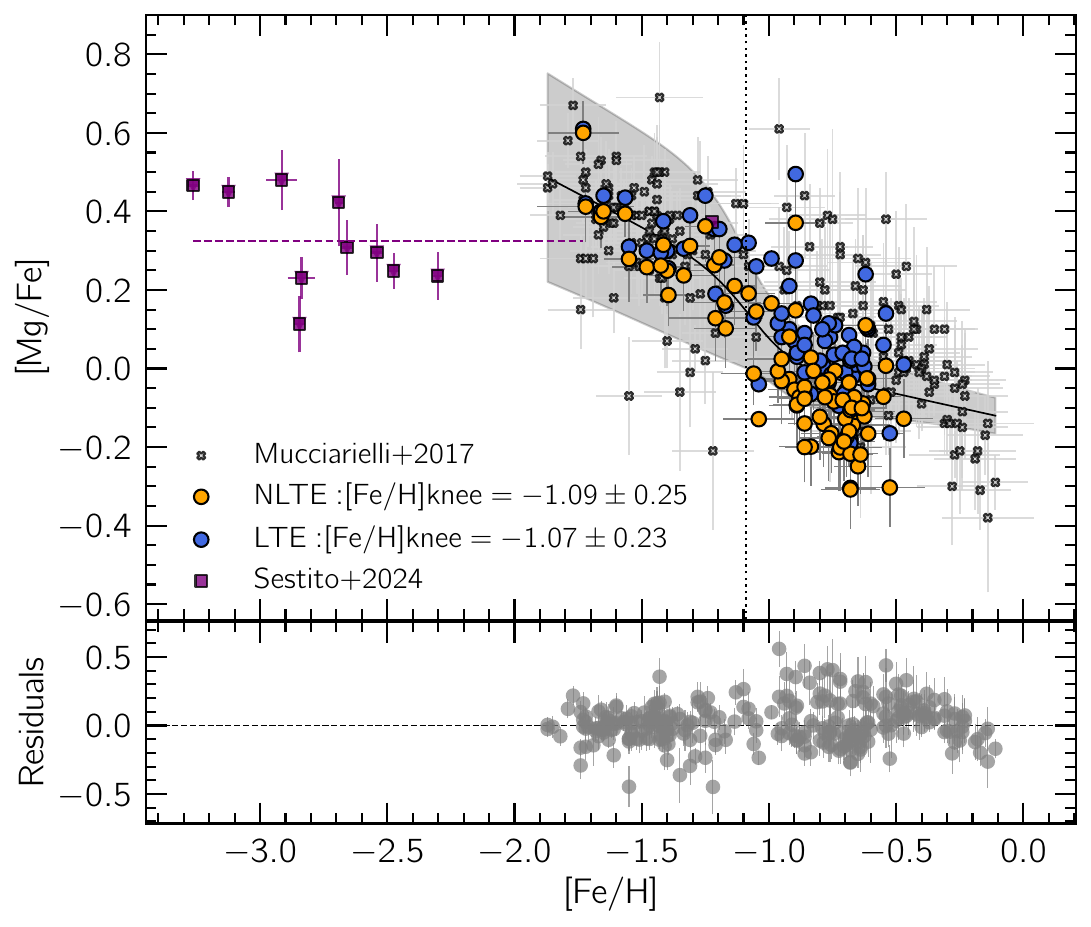} 
\caption{\small{Knee fitting using only the [Mg/Fe] abundances from our original LTE analysis (shown in blue), compared to the new NLTE-corrected values, along with two literature datasets. The dashed line marks the position of the NLTE knee ([Fe/H] = –1.09). The bottom panel reports the residual from the fitting of the $\alpha$ knee.}}
 \label{fig:knee_nlte}
\end{figure}
Figure~\ref{fig:knee_nlte} shows both the original [Mg/Fe] LTE results (in blue) and the ones corrected for NLTE effects (in orange), plus the literature measurements from \cite{2017Mucciarelli} and \cite{2024Sestito}. The two new values found for these two cases are $\rm{[Fe/H]_{LTE}} = -1.07\pm0.23$ and $\rm{[Fe/H]_{NLTE}} = -1.09\pm0.25$ which are both slightly more metal poor than the value reported in Sect.~\ref{sec:knee}, but fully comparable given the error bars.

It is not straightforward to compare the results presented here with those obtained by averaging Ca and Mg LTE measurements, as the number of stars and the literature samples included in the fitting procedure differ. However, it is noteworthy that the estimated position of the knee does not change significantly. We highlight that a shift of the knee toward more metal-poor values ([Fe/H] $\sim -1.5$ dex) cannot be ruled out, as there is a hint of a plateau in that region. Nevertheless, the lack of data around this metallicity range makes it difficult to reliably fit a knee there. This limitation could be overcome with a more homogeneous coverage in the range $-2.0<$ [Fe/H] $<-1.0$. However, in this scenario, estimating the SFH from the metallicity value of the knee depends not only on the data size, but also on the assumptions made in stellar modeling. Improving our knowledge of stellar astrophysics is thus pressing to understand further how galaxies assemble.

\section{Stellar ages}\label{sec:append_ages}

\begin{figure}[h]
\centering
 \includegraphics[width=0.34\columnwidth]{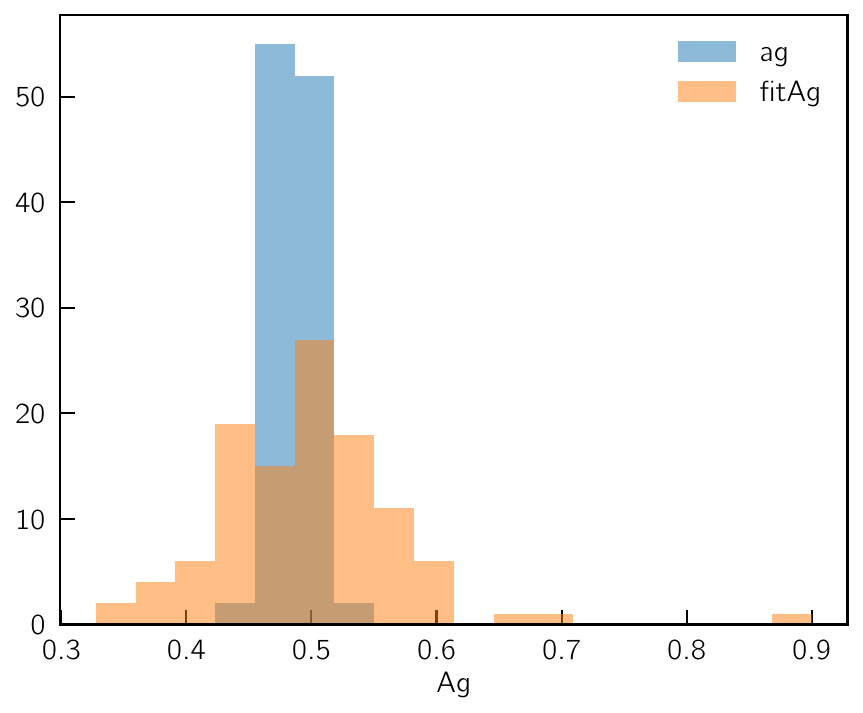} 
  \includegraphics[width=0.39\columnwidth]{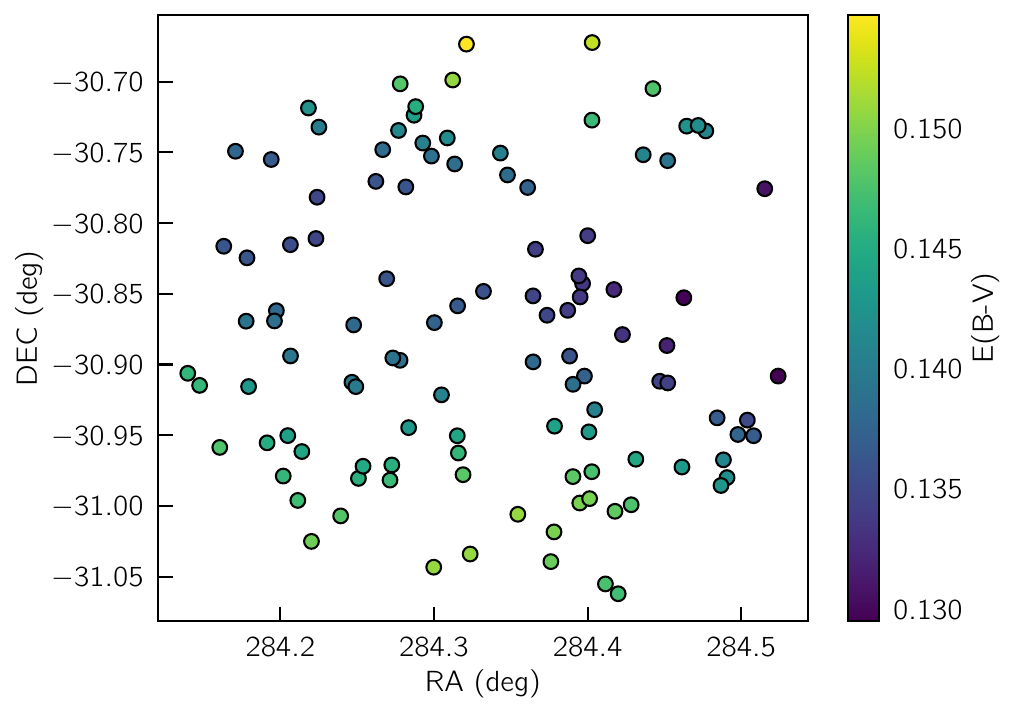}

\caption{\small{Left: Histograms of the extinction coefficient distributions derived from isochrones fitting for stellar age determination (in orange) and from the de-reddening applied for atmospheric parameter derivation (in light blue). Right: Equatorial field of the FLAMES targets, color coded by extinction values.}}
 \label{fig:exct}
\end{figure}

We compared the extinction values computed during the age derivation process with those derived for the atmospheric parameters (as explained in Sect.~\ref{sect:AP}). Specifically, we computed the extinction coefficient $A_{v} = 3.1 \times \rm{E(B-V})$ \citep[where the color index is obtained from the][dust map]{Schlegel98} and then applying the conversion $A_{g} = A_{v} * 0.85926$. This comparison is shown in the left panel of Fig.~\ref{fig:exct}. Despite a broader distribution of the coefficients fitted for inferring stellar ages (fitAg, in orange), its peak coincides with a difference of 0.1 relative to the extinction coefficient inferred for the derivation of atmospheric parameters ('ag', in light blue). The outliers with $\rm{A_{g}>0.65}$ are metal-poor stars with [Fe/H]$>-1.5$, for which isochrone fitting is known to be more challenging.

Furthermore, we examined the variation in extinction across the region covered by our targets, as shown on the right of Fig.~\ref{fig:exct}. The variation in this field is not significant, with 
E(B-V) spanning $\approx 0.2$, indicating that extinction is smooth across the target field and does not impact  the homogeneity of the derivation of atmospheric parameters.

\end{appendix}

\end{document}